\documentclass{article}
\RequirePackage{times}
\RequirePackage{mathptm}
\usepackage{graphicx,float}
\usepackage{longtable}

\newcommand{\param}[1]{\begin{small}\textsl{\,#1}\end{small}}
\newcommand{\mmexpr}[1]{\texttt{#1}}
\newcommand{\cmd}[2]{\mmexpr{#1\,[}\textnormal{#2}\mmexpr{]}\,}
\newcommand{\oparam}[1]{\begin{small}\textsl{\,opt:\,#1}\end{small}}
\newcommand{\beispiel}[3]{
  \paragraph{Example:}
  \begin{list}{bsp}{}
  \item[\mmexpr{\footnotesize{In[]=}}]{\cmd{#1}{#2}}
  \item[\mmexpr{\footnotesize{Out[]=}}]{#3}
  \end{list}}
\newcommand{\beschr}[1]{\paragraph{Description:}#1}
\newcommand{\cdd}{\texttt{cdd}}
\newcommand{\cddcmd}[1]{#1}
\newcommand{\mathematica}{\textit{Mathematica}}
\begin{document}

\title{Boole-Bell-type inequalities in \mathematica}
\author{Stefan Filipp and Karl Svozil\\
 {\small Institut f\"ur Theoretische Physik,}
  {\small Technische Universit\"at Wien }     \\
  {\small Wiedner Hauptstra\ss e 8-10/136,}
  {\small A-1040 Vienna, Austria   }            \\
  {\small e-mail: svozil@tuwien.ac.at}
 }
\date{ }
\maketitle

\begin{abstract}
Classical Pitowsky correlation polytopes are reviewed with particular emphasis
on the Minkowski-Weyl representation theorem.
The inequalities representing the faces of
polytopes are Boole's ``conditions of possible experience.''
Many of these inequalities have been discussed in the context of Bell's inequalities.
We introduce CddIF, a \mathematica\ package created as an interface between
\mathematica\ and the \cdd\ program
by Komei Fukuda, which represents a highly efficient
method to solve the hull problem
for general classical correlation polytopes.
\end{abstract}

\newpage
\tableofcontents
\newpage

\section[Boole-Bell Type Inequalities]{Boole-Bell Type Inequalities And Their Geometric Representation}
\label{sec:weg}
In the middle of the 19th century the English mathematician George Boole
formulated a theory of "conditions of possible experience"
\cite{Boole,Boole-62,Hailperin,pitowsky,Pit-94}.
These conditions are related to relative frequencies of
logically connected
events and are expressed by certain equations or inequalities.
More recently, similar equations for a particular setup which are relevant in the
quantum mechanical context have been discussed by Clauser and  Horne and others
\cite{cl-horne,chsh,clauser}.
Pitowsky has given a geometrical interpretation in terms of correlation polytopes
\cite{pitowsky-89a,pitowsky,Pit-91,Pit-94}.

   \subsection{Simple urn model}
Consider an urn containing some balls of different colors and styles:
Each ball can be described by two properties,
let us say "yellow" and "wooden",
so each ball can have the property "yellow" or the property "wooden",
but it can also have both "yellow and wooden".
The state of the urn can be given by the probabilities to draw a ball
with one of these properties: $
p_1$ is the proportion of yellow balls in the urn, $p_2$
the proportion of wooden ones and $p_{12}$ denotes
the proportion of yellow and wooden balls.
If there are enough balls in the urn these proportions are in fact the probabilities to get a ball with the special property by drawing. Clearly the inequalities
\begin{equation}
  \label{eq:propineq}
  \begin{array}{lll}
    0\leq p_{12} \leq p_{2} \leq 1 & and & 0\leq p_{12} \leq p_{1} \leq 1\\
  \end{array}
\end{equation}
are fulfilled by the proportions and so $p_1$, $p_2$ and $p_{12}$ can be seen as probabilities of two events and their joint event only if these inequalities are satisfied. Simply by taking some appropriate numbers ($p_1$ = 0.6, $p_2$=0.72 and $p_{12}$=0.32) we can see, that equations (\ref{eq:propineq}) are not sufficient. If we take the probability to draw a ball which is either yellow or wooden ($p_1$ + $p_2$ - $p_{12}$) into consideration, a new inequality can be found that is not satisfied by the numbers chosen:
\begin{equation}
  \label{eq:propineq2}
  0\leq p_1 + p_2 - p_{12} \leq 1
\end{equation}
It can be shown that the inequalities (\ref{eq:propineq}) and (\ref{eq:propineq2})
are necessary and sufficient for the numbers $p_1$, $p_2$ and $p_{12}$ to represent
probabilities of two events and their joint \cite{pitowsky}.

   \subsection{Geometrical interpretation}
\begin{figure}[htbp]
  \begin{center}
    \includegraphics[width=60mm]{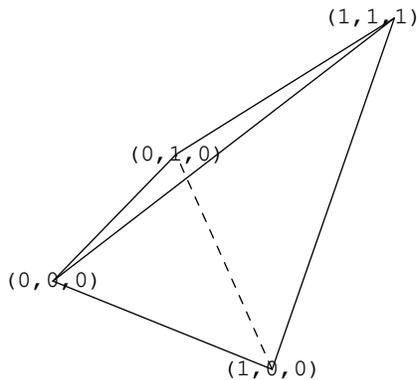}
    \caption{Polytope associtated with the  urn model}
    \label{fig:urn}
  \end{center}
\end{figure}

Itamar Pitowsky
\cite{pitowsky-89a,pitowsky,Pit-91,Pit-94} has suggested a geometric interpretation.
Consider the truth table \ref{tab:tt4a} of the above urn model,
in which $a_1$ and $a_2$  represent the statements that
``the ball drawn from the urn is yellow,''
``the ball drawn from the urn is wooden,''
and
in which $a_{12}$  represent the statement that
``the ball drawn from the urn is yellow and wooden.''
\begin{table}[htbp]
  \begin{center}
    \begin{tabular}[t]{c|c|c}
      $a_1$ & $a_2$ & $a_{12}$ \\
      \hline
      0 & 0 & 0\\
      1 & 0 & 0\\
      0 & 1 & 0\\
      1 & 1 & 1\\
    \end{tabular}
    \caption{Truth table for two propositions $a_1,a_2$ and their joint proposition $a_{12}=a_1\wedge a_2$}
    \label{tab:tt4a}
  \end{center}
\end{table}
The third ``component bit'' of the vector is a function of the first components.
Actually, the function is a multiplication, since we are dealing with
the classical logical ``and'' operation here.
Let us take the set of all numbers ($p_1$, $p_2$, $p_{12}$)
satisfying the inequalities stated above as a
set of vectors in a three-dimensional real space.
This amounts to interpreting the rows of the truth table as vectors;
the entries of the rows being the vector components.
This procedure yields a closed convex polytope with
vertices (0,0,0), (1,0,0), (0,1,0) and (1,1,1) (cf. Figure \ref{fig:urn}).
The extreme points (vertices) can be interpreted as follows:\\
(0,0,0) is a case where no yellow and no wooden balls are in the urn at all,\\
(1,0,0) is representing the configuration that all balls are yellow and no one is wooden.\\
(0,1,0) is representing the configuration that all balls are wooden and no one is yellow.\\
(1,1,1) is a case with only yellow and at the same time wooden balls.

   \subsection{Minkowski-Weyl representation theorem}

The Minkowski-Weyl representation theorem (e.g., \cite[p. 29]{ziegler}) states that  compact convex sets
are ``spanned'' by their extreme points; and furthermore that the
representation of this polytope by the inequalities corresponding to the planes of their
faces is an equivalent one.

Stated differently, every convex polytope in an Euclidean space has a dual description:
either as the convex hull of its vertices (V-representation),
or as the intersection of a finite number of half-spaces,
each one given by a linear inequality (H-representation)
This equivalence is known as the \emph{Weyl-Minkowski} theorem.

The problem to obtain all inequalities from the vertices of a convex
polytope is known as the \emph{hull problem}. One solution strategy
is the Double Description Method \cite{MRTT53} which we shall use but not review here.

   \subsection{From vertices to inequalities}

For the above simple urn model, the inequalities are rather intuitive
and can be easily obtained by guessing.
This is impossible in the general case involving more events
and more joint probabilities thereof.
In order to obtain the relevant inequalities---Boole's
``conditions of possible experience''---we have to find a hopefully
constructive way to derive them.

Recall that a vector is an element of the polytope if and only if
it can be represented as a certain bounded convex combination,
i.e., a bounded linear span,  of the vertices.
More precisely, let us denote the \emph{convex hull} $\textrm{conv}(K)$ of a finite set of points
$K=\left\{\bf{x}_1,\ldots ,\bf{x}_n\right\}\in R^d$ by
\begin{equation}
\textrm{conv} (K)=\left\{\lambda_1 \bf{x}_i+\cdots +\lambda_n \bf{x}_n \;\Big|\;
n\ge 1, \lambda_i\ge 0, \sum_{i=1}^n \lambda_i=1\right\}
.
\label{e-cs}
\end{equation}
In the probabilistic context, the coefficients $\lambda_i$ are
interpreted as the probability that the event represented
by the extreme point $\bf{x}_i$ occurs, whereby
$K$ represents the complete set of all atoms of a Boolean algebra.
The geometric interpretation of $K$ is the set of all extreme
points of the correlation polytope.

In summary, the connection between the convex hull of
the extreme points of a correlation polytope
and the inequalities representing its faces is guaranteed
by  the Minkowski-Weyl representation theorem.
A constructive solution of the corresponding hull problem exists
(but is NP-hard \cite{Pit-91}).

For the special urn model introduced above this means
that any three numbers ($p_1$, $p_2$ and $p_{12}$) must fulfill an equation
dictated by Kolmogorov's probability axioms \cite{kolmogorov2}:
 \begin{equation}
   \label{eq:convcomb}
   (p_1, p_2,p_{12})=
\lambda_1 (0,0,0) + \lambda_2 (0,1,0) + \lambda_3 (1,0,0) + \lambda_4 (1,1,1)=(\lambda_2 + \lambda_4, \lambda_3 + \lambda_4, \lambda_4).
 \end{equation}
It is important to realize that these logical possibilities are exhaustive.
By definition,
there cannot be any other classical case which is not already
included in the above possibilities $(0,0,0),(1,0,0),(0,1,0),(1,1,1)$.
Indeed, if one or more cases would be omitted, the corresponding polytope would not be
optimal; i.e., it would be embedded in the optimal one.
Therefore, any ``state'' of a physical system represented by a probability distribution
has to satisfy the constraint
\begin{equation}
  \label{eq:convconstr}
  \lambda_1 +  \lambda_2 +  \lambda_3 +  \lambda_4 = 1.
\end{equation}
The four extreme cases $\lambda_i=1, \lambda_j=0$ for $i\in \{1,2,3,4\}$ and $j\neq i$
just correspond to
the vertices spanning the classical correlation polytope as the convex sum (\ref{e-cs}).

A generalization to arbitrary configurations is straightforward.
To solve the hull problem for more general cases, an efficient algorithm has to be used.
There are some algorithms to solve this problem,
but they run in exponential time in the number of events,
thus it can be solved only for small enough cases to get a solution in conceivable time.

   \subsection{From inequalities to vertices}
Conversely, a vector is an element of the convex polytope if and only
if its coordinates satisfy a set of linear inequalities which represent the
supporting hyper-planes of that polytope.
The problem to find the extreme points (vertices) of the polytope
from the set of inequalities is dual to the hull problem considered above.

   \subsection{Quantum mechanical context}
In the quantum mechanical case the elementary irreducible
events are clicks in particle detectors
and the probabilities
have to be calculated using the formalism of quantum mechanics.
It is by no means trivial that these probabilities satisfy  Eq. (\ref{eq:convconstr}),
in particular if one realizes that quantum Hilbert lattices are nonboolean and
have an infinite number of atoms.
As it turns out, Boole's ``conditions of possible experience'' are violated if
one considers probabilities associated with complementary events,
thereby assuming counterfactuality.
(This is a development and a generalization Boole could have hardly forseen!)
\begin{figure}[!htbp]
  \begin{center}
      \includegraphics[width=80mm]{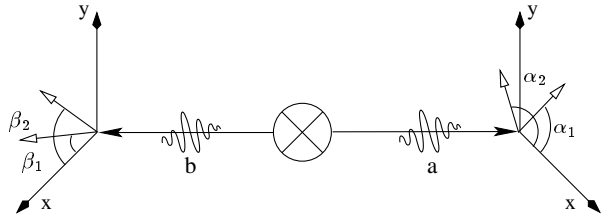}
    \caption{Experimental setting to test the violation of Boole - Bell type inequalities}
    \label{fig:22graph}
  \end{center}
\end{figure}
\par
As an example we take a source that produces pairs of spin-$\frac{1}{2}$
particles in a singlet-state
\mbox{($|\psi\rangle=\frac{1}{\sqrt{2}}(|\uparrow\downarrow\rangle-
|\downarrow\uparrow\rangle)$)}.
The particles fly apart along the z axis and after the particles have separated,
measurements on spin components along one out of two directions are made.
If, for simplicity, the measurements are made in the x-y plane perpendicular
to the trajectory of the particles, the direction of the measurement can be given
by angles measured from the vertical x axis ($\alpha_1$ and $\alpha_2$ on the one side, $\beta_1$ and $\beta_2$ on the other side). On each side the measurement angle is chosen randomly for each pair of incoming particles and each measurement can yield two results - in $\frac{\hbar}{2}$ units: ``+1'' for spin up and ``-1''
for spin down (cf. Figure \ref{fig:22graph}).

Deploying this configuration we get probabilities to find a particle measured
along the axis specified by the angles $\alpha_1$, $\alpha_2$, $\beta_1$ and $\beta_2$
either in spin up or in spin down state  denoted as $p_{a1}$, $p_{a2}$, $p_{b1}$, $p_{b2}$
-  and we also take the joint event of finding a particle on one side at the
angle $\alpha_1$ ($\alpha_2$)
in a specific spin state and the other particle on the other side
along the vector $\beta_1$ ($\beta_2$) in a specific spin state,
 denoted as $p_{a1b1}$, $p_{a2b1}$, $p_{a1b2}$ and $p_{a2b2}$.
To construct the convex polytope to this experiment
we build up a truth table of all possible events using a ``1'' as ``spin up is detected along the specific axis'' and a ``0'' as ``spin down is detected along the specific axis'' (table \ref{tab:tt4}).
\begin{table}[htbp]
  \begin{center}
    \begin{tabular}[t]{c|c|c|c|c|c|c|c}
      $\alpha_1$ & $\alpha_2$ & $\beta_1$ & $\beta_1$ &
      $\alpha_1 \beta_1$ & $\alpha_1 \beta_2$ & $\alpha_2 \beta_1$ & $\alpha_2 \beta_2$\\
      \hline
      0 & 0 & 0 & 0 & 0 & 0 & 0 & 0\\
      1 & 0 & 0 & 0 & 0 & 0 & 0 & 0\\
      0 & 1 & 0 & 0 & 0 & 0 & 0 & 0\\
      1 & 1 & 0 & 0 & 0 & 0 & 0 & 0\\
      0 & 0 & 1 & 0 & 0 & 0 & 0 & 0\\
      1 & 0 & 1 & 0 & 1 & 0 & 0 & 0\\
      0 & 1 & 1 & 0 & 0 & 0 & 1 & 0\\
      1 & 1 & 1 & 0 & 1 & 0 & 1 & 0\\
      0 & 0 & 0 & 1 & 0 & 0 & 0 & 0\\
      1 & 0 & 0 & 1 & 0 & 1 & 0 & 0\\
      0 & 1 & 0 & 1 & 0 & 0 & 0 & 1\\
      1 & 1 & 0 & 1 & 0 & 1 & 0 & 1\\
      0 & 0 & 1 & 1 & 0 & 0 & 0 & 0\\
      1 & 0 & 1 & 1 & 1 & 1 & 0 & 0\\
      0 & 1 & 1 & 1 & 0 & 0 & 1 & 1\\
      1 & 1 & 1 & 1 & 1 & 1 & 1 & 1\\
    \end{tabular}
    \caption{Truth table for four propositions}
    \label{tab:tt4}
  \end{center}
\end{table}
The rows of this table are then identified with the vertices of the convex polytope.
By using the Minkowski-Weyl theorem and by
solving the hull problem,
the vertices determine
the hyper-planes confining the polytope,
 i.e. the inequalities which the probabilities have to satisfy.
As a result the following inequalities are gained:
\begin{equation}
  \label{eq:chineq}
  \begin{array}{ll}
    0\leq p_{aibi}\leq p_{ai}\leq 1, 0\leq p_{aibi}\leq p_{bi}\leq 1 & i = 1,2\\
    p_{ai} + p_{bi} - p_{aibi} \leq 1 & i = 1,2\\
  \end{array}
\end{equation}
\begin{equation}
  \label{eq:chineq2}
  \begin{array}{l}
    -1\leq p_{a1b1} + p_{a 1b2} + p_{a2b2} - p_{a2b1} - p_{a1} - p_{b2} \leq 0 \\
    -1\leq p_{a2b1} + p_{a2b2} + p_{a1b2} - p_{a1b1} - p_{a2} - p_{b2} \leq 0 \\
    -1\leq p_{a1b2} + p_{a1b1} + p_{a2b1} - p_{a2b2} - p_{a1} - p_{b1} \leq 0 \\
    -1\leq p_{a2b2} + p_{a2b1} + p_{a1b1} - p_{a1b2} - p_{a2} - p_{b1} \leq 0 \\
  \end{array}
\end{equation}

The last four inequalities are known as \emph{Clauser-Horne inequalities}. As noticed above the probabilities have to be seen in a quantum mechanical context. If we consider the singlet state of spin-$\frac{1}{2}$ particles $|\psi\rangle=\frac{1}{\sqrt{2}}(|\uparrow\downarrow\rangle-|\downarrow\uparrow\rangle)$ it is well known that the probability to find the particles both either in spin up or in spin down states is given by $P^{\uparrow\uparrow}(\theta)=P^{\downarrow\downarrow}(\theta)=\frac{1}{2}sin^2(\theta/2)$ - where $\theta$ is the angle between the measurement directions. The single event probability is clearly $p_i = \frac{1}{2}$. By choosing
\begin{equation}
  \label{eq:angles}
  a_1 = -\frac{\pi}{3}\qquad a_2 = b_1 = \frac{\pi}{3}\qquad b_2 = \frac{\pi}{3}
\end{equation}
as measurement directions, we get for $p=(p_{a1},p_{a2},p_{b1},p_{b2},p_{a1b1},p_{a2b1},p_{a1b2},p_{a2b2})$:
\begin{equation}
  \label{eq:chprobsol}
  p=(\frac{1}{2},\frac{1}{2},\frac{1}{2},\frac{1}{2},\frac{3}{8},\frac{3}{8},0,\frac{3}{8})
\end{equation}
and one of the inequalities found in (\ref{eq:chineq2}) is violated:
\begin{equation}
  \label{eq:violch}
  p_{a1b1} + p_{a1b2} + p_{a2b2} - p_{a2b1} - p_{a1} - p_{b2} = \frac{3}{8} + \frac{3}{8} + \frac{3}{8} - 0 - \frac{1}{2} - \frac{1}{2} = \frac{1}{8} > 0
\end{equation}

The generalization is straightforward.
Violations of certain inequalities involving classical probabilisties---Boole's
``conditions of possible experience'' \cite{Boole-62}---also appear in higher dimensions in
configurations containing more particles and/or more measurement directions.
We shall consider more examples below.

\section{Installation}
\label{sec:install}

\subsection{\mathematica}
\label{sec:inmm}
All functions described in the following section can be found in
the \mathematica-package \textit{cddif.m}.
In general this package has to be loaded into the
current \mathematica-kernel by the command
\mmexpr{<<{\small'path to cddif.m'}/cddif.m}, short
description and usage of the functions is available by entering \mmexpr{?{\small'<function>'}}.

To guarantee a proper run of all functions it is necessary (and hopefully sufficient)
that \cdd\ is located in any directory listed in the PATH-variable (usually /bin, /usr/bin, /usr/local/bin, \ldots)\footnote{for setting environment variables look at the manual of \textit{set} or \textit{env}} or in the current working directory, which can be shown by evaluating \mmexpr{Directory[]} or changed using the function \cmd{SetDirectory}{\param{directory\_String}}. If one likes to avoid the frequent use of this function one can append this command to the package-file \textit{cddif.m} before the line \mmexpr{End[]} so that on each loading of the package the directory is set automatically to a personal working directory.

\subsection{cdd}
\label{sec:incdd}

\cdd\ is a C++ (ANSI C) implementation of the Double Description Method
\cite{MRTT53} by Komei Fukuda\cite{cdd-pck}.
It generates all vertices (i.e. extreme points)
and extreme rays of a general convex polyhedron given by a system of
linear inequalities.
Conversely, it solves the hull
problem by generating a system of linear inequalities given all vertices.

At this point we refer to the documentation of the program for
the installation of the \cdd\ - package, in particular to the file \textit{cdd.readme}
included in the package and to the online documentation at
http://www.ifor.math.ethz.ch/\~{}fukuda/cddman/cddman.html. \cdd\
is available for free and one can download it from the homepage of
Komei Fukuda\cite{cdd-pck} (\textit{http://www.ifor.math.ethz.ch/\~{}fukuda/cdd\_home/cdd.html}), where one can also find a manual to the usage of \cdd, especially descriptions to the format of the input- and output-files and to options that can be passed to \cdd.

\subsection{Installation on Windows-platforms}
\label{sec:inwin}
Currently a version of \cdd\  executable on Windows platforms can be downloaded from
http://tph.tuwien.ac.at/\~{}svozil/cdd/cdd.exe
(a different compilation is at http://www.wis.kuleuven.ac.be/wis/algebra/kathleen/files/cdd061.exe).
\par\noindent
On Windows-systems \mathematica\ must be able to find \cdd\ in a directory listed in the PATH-variable or in the current working directory, too. To set the PATH-variable in Windows 2000/NT go to the ``control panel'' and click on the ``system properties'', then click ``advanced'' and there is a place where the variable PATH is specified. Here one can add the path to \cdd\ (separated by a semicolon). In WindowsME one needs to go execute ``msconfig'' to get to the System Configuration Utility - in ``Environment'' one can edit the PATH-variable and for Windows98/95 one must edit the file \textit{``autoexec.bat''} to get the path set.
\par\noindent
Finally it can be necessary to rename the \cdd\ - executable file (e.g. from \textit{cdd061.exe}) to \textit{cdd.exe}. The CddIF-Package uses \textit{cdd} as default command to run \cdd, using the function \cmd{SetCddCmd}{\param{cmd\_String}} one can change this behavior. Like already stated above one can also add this commandline to the package-file \textit{cddif.m} just before the \mmexpr{End[]}-statement to change the default command automatically when loading the package.

\section{Description Of Functions}
\label{sec:func}

In this section all functions of the \emph{CddIF} - package are listed. For each function the syntax including the necessary parameters (if parameters are optional, it has an ``\oparam{}'' as prefix), a description and an example is given.

\subsection{CddFormat}
\label{sec:cddformat}

\cddcmd{\cmd{CddFormat}{\param{vertices},\oparam{options}}}
  \begin{list}{param}{}
  \item[\param{vertices (List):}]List of m vertices in n dimensions of the form
    \par\noindent\mmexpr{\{\{$x_{11}$,$x_{12}$,\ldots,$x_{1n}$\},\{$x_{21}$,\ldots,$x_{2n}$\},\ldots,\{$x_{m1}$,$x_{m2}$,\ldots,$x_{mn}$\}\}}
  \item[\param{options (List):}]Options to \cdd\ (e.\,g. \emph{adjacency, nondegenerate, minindex,...}) - see documentation to \cdd\ (http://www.ifor.math.ethz.ch/~fukuda/cddman/cddman.html)
  \end{list}
\beschr{A list of vertices, which can be determined for example by \cmd{TruthTable}{...,\param{IncludeVars$\rightarrow$False}}, are converted to a format recognized by \cdd. Additionally options to \cdd\ can be declared.}
\beispiel{CddFormat}{\param{\{\{1,0,0\},\{0,1,0\},\{1,1,1\}\}}}{\mmexpr{\{V-representation begin,\{3, 4, integer\},\\\{1, 1, 0, 0\}, \{1, 0, 1, 0\}, \{1,1, 1, 1\},\\end\}}}

\subsection{ToCddExtFile}
\label{sec:tocddextfile}

\cddcmd{\cmd{ToCddExtFile}{\param{file},\param{vertices},\oparam{options}} or
  \\\cmd{ToCddExtFile}{\param{file},\param{particles},\param{measurements},\oparam{options}}}
\begin{list}{param}{}
\item[\param{file (String):}]Filename for output of H-representation (\emph{``.ext''}-suffix is automatically appended)
\item[\param{vertices (List):}]List of m vertices in n dimensions of the form
  \par\noindent \mmexpr{\{\{$x_{11}$,$x_{12}$,\ldots,$x_{1n}$\},\{$x_{21}$,\ldots,$x_{2n}$\},\ldots,\{$x_{m1}$,$x_{m2}$,\ldots,$x_{mn}$\}\}}
\item[\param{options (List):}]Options to \cdd\ (e.\,g. \emph{adjacency, nondegenerate, minindex,...}) - see documentation to \cdd\ (http://www.ifor.math.ethz.ch/~fukuda/cddman/cddman.html)
\item[\param{particles (Integer):}]Number of particles
\item[\param{measurements (Integer):}]Number of possible measurements to each particle (equivalent to number of detection angles)
\end{list}
\beschr{Creates a file with \emph{``.ext''}-extension that contains the data of the given configuration to use in \cdd. Eighter a list of vertices of the considered correlation polytop or the number of particles used and the possible measurements to each can be handed over. In the latter case the list of vertices is generated automatically.}
\beispiel{ToCddExtFile}{\param{``test'',2,3}}{\mmexpr{test.ext}\\writes the file \emph{``test.ext''} to the current working directory, containing the vertices of the 2-particles 3-measurements configuration.}

\subsection{TruthTable}
\label{sec:truthtable}

\cddcmd{\cmd{TruthTable}{\param{particles},\param{measurements},\oparam{options}}}
  \begin{list}{param}{}
  \item[\param{particles (Integer):}]Number of particles
  \item[\param{measurements (Integer):}]Number of possible measurements to each particle (equivalent to number of detection angles)
\item[\oparam{options:}]The only possible option is \mmexpr{IncludeVars $\rightarrow$ True/False}. If \mmexpr{IncludeVars $\rightarrow$ True} is defined, the function includes a list of variables belonging to the given configuration as titles of the columns and output will be made in \mmexpr{MatrixForm}, otherwise a list containing all vertices is returned. Default is \mmexpr{IncludeVars $\rightarrow$ True}.
  \end{list}
\beschr{Creates a truth table of the given configuration, containing all vertices of the corresponding correlation polytopes. For generating this table all possible single events are rated eighter 0 or 1 (i. e. true or false) and the joint events are evaluated using the logical \emph{AND} operation.}
\beispiel{TruthTable}{\param{2,2}}{
  \begin{displaymath}
    \begin{array}[p]{cccccccc}

    a1 & a2 & b1 & b2 & a1b1 & a1b2 & a2b1 & a2b2\\
    0 & 0 & 0 & 0 & 0 & 0 & 0 & 0\\
    1 & 0 & 0 & 0 & 0 & 0 & 0 & 0\\
    0 & 1 & 0 & 0 & 0 & 0 & 0 & 0\\
    1 & 1 & 0 & 0 & 0 & 0 & 0 & 0\\
    0 & 0 & 1 & 0 & 0 & 0 & 0 & 0\\
    1 & 0 & 1 & 0 & 1 & 0 & 0 & 0\\
    0 & 1 & 1 & 0 & 0 & 0 & 1 & 0\\
    1 & 1 & 1 & 0 & 1 & 0 & 1 & 0\\
    0 & 0 & 0 & 1 & 0 & 0 & 0 & 0\\
    1 & 0 & 0 & 1 & 0 & 1 & 0 & 0\\
    0 & 1 & 0 & 1 & 0 & 0 & 0 & 1\\
    1 & 1 & 0 & 1 & 0 & 1 & 0 & 1\\
    0 & 0 & 1 & 1 & 0 & 0 & 0 & 0\\
    1 & 0 & 1 & 1 & 1 & 1 & 0 & 0\\
    0 & 1 & 1 & 1 & 0 & 0 & 1 & 1\\
    1 & 1 & 1 & 1 & 1 & 1 & 1 & 1

    \end{array}
  \end{displaymath}}

\subsection{RunCdd}
\label{sec:runcdd}

\cddcmd{\cmd{RunCdd}{\param{file}}}
  \begin{list}{param}{}
  \item[\param{file (String):}]File handed over to \cdd\ as command parameter (automatically extended with \emph{``ext.''}-suffix.
  \end{list}
\beschr{\cmd{RunCdd}{...} executes \cdd\ (with file as parameter) and returns the corresponding H-representation, which can be used in various other functions like \cmd{PlotInequalities}{\ldots} or \cmd{GetViolInequalities}{\ldots}.
\par\noindent Using this function one has to pay attention to the potentially long runtime in calculating the faces (i.\,e. the inequalities) of the correlation polytope. It can be more beneficial to use \cmd{ToCddExtFile}{\param{file,...}} to create a \emph{``.ext''}-file, followed by executing \cdd\ outside of \mathematica\ (eventually on a faster computer) to convert the date to H-representation stored in an \emph{``.ine''}-file. Afterwards one can read in this file utilizing \cmd{ReadInHRep}{\param{file}}}
\beispiel{RunCdd}{\param{``test''}}{\mmexpr{\{\{H-representation\},\{begin\},\{684,16,real\},\{2,0,-2,....\},....,\{end\}\}},\\whereas 2-particles 3-measurement configuration is taken into consideration here.}

\subsection{ShowVRep}
\label{sec:showvrep}

\cddcmd{\cmd{ShowVRep}{\param{particles},\param{measurements}}}
  \begin{list}{param}{}
  \item[\param{particles (Integer):}]Number of particles
  \item[\param{measurements (Integer):}]Number of possible measurements to each particle (equivalent to number of detection angles)
  \end{list}
\beschr{Shows the V-representation of a given configuration.}
\beispiel{ShowVRep}{\param{2,2}}{\mmexpr{\{V-representation\\begin,\\\{16, 9, integer\},\\\{1, 0, 0, 0, 0, 0, 0, 0,0\},\{1, 1, 0, 0, 0, 0, 0, 0, 0\},\\\{1, 0, 1, 0, 0, 0, 0, 0, 0\},\{1, 1, 1, 0, 0, 0, 0, 0, 0\},\\\{1, 0, 0, 1, 0, 0, 0, 0, 0\},\{1, 1, 0, 1, 0, 1, 0, 0, 0\},\\\{1, 0, 1, 1, 0, 0, 0, 1, 0\},\{1, 1, 1, 1, 0, 1, 0, 1, 0\},\\\{1, 0, 0, 0, 1, 0, 0, 0, 0\},\{1, 1, 0, 0, 1, 0, 1, 0, 0\},\\\{1, 0, 1, 0, 1, 0, 0, 0, 1\},\{1, 1, 1, 0, 1, 0, 1, 0, 1\},\\\{1, 0, 0, 1, 1, 0, 0, 0, 0\},\{1, 1, 0, 1, 1, 1, 1, 0, 0\},\\\{1, 0, 1, 1, 1, 0, 0, 1, 1\},\{1, 1, 1, 1, 1, 1, 1, 1, 1\},\\end\}}}

\subsection{ConvToHRep}
\label{sec:convtohrep}

\cddcmd{\cmd{ConvToHRep}{\param{particles},\param{measurements},\oparam{file},\oparam{options}} or
  \\\cmd{ConvToHRep}{\param{vertices},\oparam{file},\oparam{options}}}
  \begin{list}{param}{}
  \item[\param{particles (Integer):}]Number of particles
  \item[\param{measurements (Integer):}]Number of possible measurements to each particle (equivalent to number of detection angles)
  \item[\param{vertices (List):}]List of m vertices in n dimensions of the form
    \par\noindent\mmexpr{\{\{$x_{11}$,$x_{12}$,\ldots,$x_{1n}$\},\{$x_{21}$,\ldots,$x_{2n}$\},\ldots,\{$x_{m1}$,$x_{m2}$,\ldots,$x_{mn}$\}\}}
  \item[\param{file (String):}]Filename that is used for the conversion from a \emph{``.ext''}-file to a \emph{``.ine''}-file which is equivalent to a conversion from V-representation to H-representation. Default is \emph{``tmp''}.
  \item[\param{options (List):}]Options to \cdd\ (e.\,g. \emph{adjacency, nondegenerate, minindex,...}) - see documentation to \cdd(http://www.ifor.math.ethz.ch/~fukuda/cddman/cddman.html)
  \end{list}
\beschr{This function converts a given configuration (n particles, m measurements)
or a given list of vertices from V-representation into a H-representation.
As above in (\ref{sec:runcdd}) the potentially long calculation time has to be taken into consideration, depending on the complexity of the problem.}
\beispiel{ConvToHRep}{\param{2,3,''2\_3''}}{\mmexpr{\{\{H-representation\},\{begin\},\{684, 16,real\},\{2, 0,...\}...,\{end\}\}},\\wheras in this case the files \emph{``2\_3.ext''} (created by \mathematica\ containing the data for \cdd) and \emph{``2\_3.ine''} (created by \cdd\ as result of the calculation) are generated in the current working directory.}

\subsection{ReadInHRep}
\label{sec:readinhrep}

\cddcmd{\cmd{ReadInHRep}{\param{file}}}
  \begin{list}{param}{}
  \item[\param{file (String):}]\emph{''.ine''} - file containing the H-representation which is to be read in.
  \end{list}
\beschr{Reads the H-representation from a given \emph{``.ine''}-file for further use in various functions like \cmd{GetViolInequalities}{\ldots} or \cmd{PlotInequalities}{\ldots}.}
\beispiel{ReadInHRep}{\param{``2\_3''}}{\{\{H-representation\},\{begin\},\{684, 16,real\},\{2, 0,...\}...,\{end\}\}}

\subsection{GetInequFromHRep}
\label{sec:getinequfromhrep}

\cddcmd{\cmd{GetInequFromHRep}{\param{hrep}}}
  \begin{list}{param}{}
  \item[\param{hrep (List):}]H-representation yielded for example from the function \cmd{ReadInHRep}{\ldots} or \cmd{ConvToHRep}{\ldots}.
  \end{list}
\beschr{Returns the inequalities from a given H-representation as a list. To make the list more readable one can apply \cmd{InequToRead}{\ldots} on it.}
\beispiel{GetInequFromHRep}{\cmd{ConvToHRep}{2,1}}{\mmexpr{\{\{a1 - a1b1 + b1, 1\}, \{-a1 + a1b1, 0\}, \{a1b1 - b1, 0\}, \{-a1b1, 0\}\}}}

\subsection{InequToRead}
\label{sec:inequtoread}

\cddcmd{\cmd{InequToRead}{\param{inequalities}}}
  \begin{list}{param}{}
  \item[\param{inequalites (List):}]List of inequalities yielded from \cmd{GetInequFromHRep}{\ldots}
  \end{list}
\beschr{Makes the list of inequalities yielded from \cmd{GetInequFromHRep}{\ldots} more readable.}
\beispiel{InequToRead}{\cmd{GetInequFromHRep}{\cmd{ConvToHRep}{2,1}}}{
  \begin{math}
    \begin{array}[t]{c}
      a1-a1b1+b1\leq1\\
      -a1+a1b1\leq0\\
      a1b1-b1\leq0\\
      -a1b1\leq0
    \end{array}
  \end{math}}

\subsection{GetViolInequalities}
\label{sec:getviolinequalities}

\cddcmd{\cmd{GetViolInequalities}{\param{hrep},\param{angles},\param{functions},\param{inequ-nr},\param{violation}}} or
\par\noindent
\cddcmd{\cmd{GetViolInequalities}{\param{file},\param{angles},\param{functions},\param{inequ-nr},\param{violation},\oparam{options}}}
  \begin{list}{param}{}
  \item[\param{hrep (List):}]H-representation yielded for example from the function \cmd{ReadInHRep}{\ldots} or \cmd{ConvToHRep}{\ldots}.
  \item[\param{file (String):}]File containing the demanded H-representation in \cdd\ format.
  \item[\param{angles (List):}]List of measurement angles for each particle, whereas the dimension of the list must represent the configuration. If one chooses the configuration ``2 particles - 2 measurements'' the list must have the dimension \mmexpr{[2,2]}, in this case the particles a and b are measured each along two axis given by the angles $a_1$, $a_2$, $b_1$ and $b_2$, so this parameter has the form \mmexpr{\{\{$a_1$\,,\,$a_2$\}\,,\,\{$b_1$\,,\,$b_2$\}\}.}
  \item[\param{functions (Symbol):}]Functions to calculate the quantum mechanical probability of the events. Considering for example two spin-$\frac{1}{2}$ particles in a singlett state, the probability to find the particles both either in spin ``up'' or both in spin ``down'' states is given by $\mathbf{P}^{\uparrow\uparrow}(\theta,\phi)=\mathbf{P}^{\downarrow\downarrow}(\theta,\phi)=\frac{1}{2}\sin^2[\frac{(\theta - \phi)}{2}]$, where $\theta$ and $\phi$ are the measurement angles of the particles.\\In defining these functions one has to notice, that for all possible events (single events, two-particle-events,\ldots) an apropriate function definition has to exist, each taking a list as parameter (e.g. \mmexpr{Prob[\{x\_,y\_\}]=$\frac{1}{2}$Sin[(x\,-\,y)/2]$^2$}).
  \item[\param{inequ-nr (List):}]Range of rows in H-representation used for checking violated inequalities. Specifying this can be useful, if many inequalities have to be evaluated. The form of the parameter is \mmexpr{\{min,max\}} oder \mmexpr{All}.
  \item[\param{violation (Real):}]Only inequalities are printed out, that are violated more than this parameter. Default value is 0.
  \item[\oparam{options:}]Options can be \mmexpr{PrintOut $\rightarrow$ True/False}, which specifies, if the inequalities are printed out during evaluation or not.
  \end{list}
\beschr{\cmd{GetViolInequalities}{\ldots} calculates the discrepancy of inequalities using the given probability functions and therefore the quantum mechanical violation for a distinct adjustment (i. e. special angles) of the detectors and returns all violated inequalities.}
\beispiel{GetViolInequalities}{\cmd{ConvToHRep}{2,2},\{\{-$\frac{\pi}{6}$,0\},\{0,$\frac{\pi}{6}$\}\},Prob}{\mmexpr{(-a1b1+a1b2+a2-a2b1-a2b2+b1$\leq$1\ $\frac{9}{8}$)}\\The probability for the single event of measuring one particle in spin ``up'' or spin ``down'' at any angle is given by \mmexpr{Prob[\{x\_\}]=$\frac{1}{2}$} and the probability of the joint event has been calculated by \mmexpr{Prob[\{x\_,y\_\}]=$\frac{1}{2}$Sin[(x\,-\,y)/2]$^2$}.}

\subsection{PlotInequalities}
\label{sec:plotinequalities}

\cddcmd{\cmd{PlotInequalities}{\param{hrep},\param{range},\param{angles},\param{functions},\oparam{options}}} or
\par\noindent
\cddcmd{\cmd{PlotInequalities}{\param{file},\param{range},\param{angles},\param{functions},\param{inequ-nr},\param{violation},\oparam{options}}}

  \begin{list}{param}{}
  \item[\param{hrep (List):}]H-representation yielded for example from the function \cmd{ReadInHRep}{\ldots} or \cmd{ConvToHRep}{\ldots}.
  \item[\param{file (String):}]File containing the demanded H-representation in \cdd\ format.
  \item[\param{range (List):}]Parameter specifying the range for the variable x, which is plotted on the horizontal axis. It has the form \{x\,,\,x$_{min}$\,,\,x$_{max}$\}(see \mathematica\ function \cmd{Plot}{\ldots})
  \item[\param{angles (List):}]List of measurement angles for each particle, whereas the dimension of the list must represent the configuration. If one chooses the configuration ``2 particles - 2 measurements'' the list must have the dimension \mmexpr{[2,2]}, in this case the particles a and b are measured each along two axis given by the angles $a_1$, $a_2$, $b_1$ and $b_2$, so this parameter has the form \mmexpr{\{\{$a_1$\,,\,$a_2$\}\,,\,\{$b_1$\,,\,$b_2$\}\}.}
 \item[\param{functions (Symbol):}]Functions to calculate the quantum mechanical probability of the events. Considering for example two spin-$\frac{1}{2}$ particles in a singlett state, the probability to find the particles both either in spin ``up'' or both in spin ``down'' states is given by $\mathbf{P}^{\uparrow\uparrow}(\theta,\phi)=\mathbf{P}^{\downarrow\downarrow}(\theta,\phi)=\frac{1}{2}\sin^2[\frac{(\theta - \phi)}{2}]$, where $\theta$ and $\phi$ are the measurement angles of the particles.\\In defining these functions one has to notice, that for all possible events (single events, two-particle-events,\ldots) an apropriate function definition has to exist, each taking a list as parameter (e.g. \mmexpr{Prob[\{x\_,y\_\}]=$\frac{1}{2}$Sin[(x\,-\,y)/2]$^2$}).
 \item[\param{inequ-nr (List):}]Range of rows in H-representation used for checking violated inequalities. Specifying this can be useful, if many inequalities have to be evaluated. The form of the parameter is \mmexpr{\{min,max\}} oder \mmexpr{All}.
 \item[\param{violation (Real):}]Only inequalities are printed out, that are violated more than this parameter. Default value is 0.
 \item[\oparam{options:}]Options for the \mathematica - function \cmd{Plot}{\ldots} can be handed over.
  \end{list}
\beschr{This function yields a plot of (violated) inequalities, whereas the function plotted is $f(x)=p(x)-b$ derived from the inequalites of the form $p(x) \leq b$ ($p(x)$ is a linear combination of functions to calculate the probabilities of (joint) events, dependent on one variable). Consequently the degree of violation is represented as a positive value of $f(x)$.
\par
Take for example the case ``2 particles and 2 measurement directions'', where the inequality
  \begin{displaymath}
    -p_{a1b1}+p_{a1b2}+p_{a2}-p_{a2b1}-p_{a2b2}+p_{b1} \leq 1
  \end{displaymath}
appears. The probability for the single event of measuring one particle in spin ``up'' or spin ``down'' at any angle is given by $$p_{a1}(x)=p_{a2}(x)=p_{b1}(x)=p_{b2}(x)=\frac{1}{2}$$ and the probability of the joint event can be calculated by $$p_{a1b1}(x,y)=p_{a2b1}(x,y)=p_{a1b1}(x,y)=p_{a2b2}(x,y)=\frac{1}{2}\sin[(x\,-\,y)/2]^2.$$ If we define the measurement angles by $$a_1 = -\frac{\pi}{3}+x\qquad a_2 = b_1 = 0\qquad b_2 = 2*\pi$$ the inequality can be written as
\begin{displaymath}
  1+\frac{1}{2}\sin{\frac{1}{2}(-\frac{\pi}{3}-x)}²-\frac{\sin{x}²}{2}-\frac{1}{2}\sin{\frac{1}{2}(-\frac{\pi}{3}+x)}² \leq 1
\end{displaymath}
The left side is dependent on x ($p(x) = 1+\frac{1}{2}\sin{\frac{1}{2}(-\frac{\pi}{3}-x)}²-\frac{\sin{x}²}{2}-\frac{1}{2}\sin{\frac{1}{2}(-\frac{\pi}{3}+x)}²$) and $b = 1$.
The function to be plotted is $f(x)=p(x)-b$, thus
\begin{displaymath}
   f(x) = 1+\frac{1}{2}\sin{\frac{1}{2}(-\frac{\pi}{3}-x)}²-\frac{\sin{x}²}{2}-\frac{1}{2}\sin{\frac{1}{2}(-\frac{\pi}{3}+x)}² - 1
\end{displaymath}
}
\beispiel{PlotInequalities}{\cmd{ConvToHRep}{2,2},\{$x$,0,$\pi$\},\{\{-$\frac{\pi}{3},+x$,0\},\{0,$2\,x$\}\},Prob}{\
      \begin{figure}[H]
        \begin{center}
          \includegraphics[width=70mm]{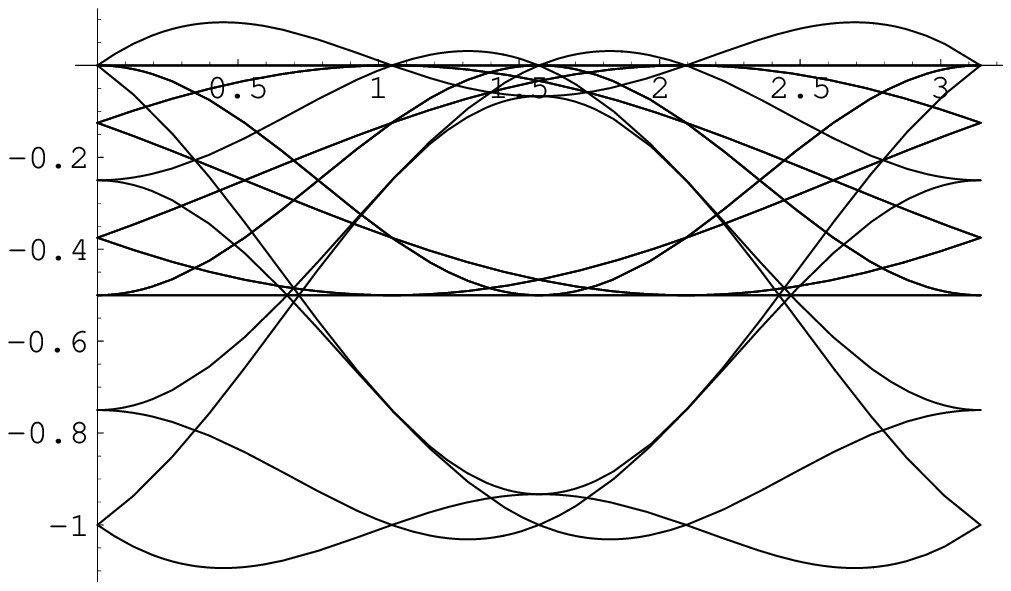}
          \label{fig:22plot}
        \end{center}
      \end{figure}
The functions to calculate the probabilities have been defined as \mmexpr{Prob[\{x\_\}]:=$\frac{1}{2}$} for a single event and \mmexpr{Prob[\{x\_,y\_\}]:=$\frac{1}{2}$Sin[(x\,-\,y)/2]$^2$} for two-particle events.}

\subsection{ContPlotInequalities}
\label{sec:contplot}

\cddcmd{\cmd{ContPlotInequalities}{\param{hrep},\param{rangex},\param{rangey},\param{angles},\param{func},\param{ineq-nr},\param{violation},\oparam{options}} or
\par\noindent
\\\cmd{ContPlotInequalities}{\param{file},\param{rangex},\param{rangey},\param{angles},\param{func},\param{ineq-nr},\param{violation},\oparam{options}}}
  \begin{list}{param}{}
  \item[\param{hrep (List):}]H-representation yielded for example from the function \cmd{ReadInHRep}{\ldots} or \cmd{ConvToHRep}{\ldots}.
  \item[\param{rangex (List):}]Parameter specifying the range for the variable x, which is plotted on the horizontal axis. It has the form \{x\,,\,x$_{min}$\,,\,x$_{max}$\}(see \mathematica\ function \cmd{Plot}{\ldots})
  \item[\param{rangex (List):}]Parameter specifying the range for the variable y, which is plotted on the vertical axis. It has the form \{y\,,\,y$_{min}$\,,\,y$_{max}$\}(see \mathematica\ function \cmd{Plot}{\ldots})
  \item[\param{angles (List):}]List of measurement angles for each particle, whereas the dimension of the list must represent the configuration. If one chooses the configuration ``2 particles - 2 measurements'' the list must have the dimension \mmexpr{[2,2]}, in this case the particles a and b are measured each along two axis given by the angles $a_1$, $a_2$, $b_1$ and $b_2$, so this parameter has the form \mmexpr{\{\{$a_1$\,,\,$a_2$\}\,,\,\{$b_1$\,,\,$b_2$\}\}.}
 \item[\param{functions (Symbol):}]Functions to calculate the quantum mechanical probability of the events. Considering for example two spin-$\frac{1}{2}$ particles in a singlett state, the probability to find the particles both either in spin ``up'' or both in spin ``down'' states is given by $\mathbf{P}^{\uparrow\uparrow}(\theta,\phi)=\mathbf{P}^{\downarrow\downarrow}(\theta,\phi)=\frac{1}{2}\sin^2[\frac{(\theta - \phi)}{2}]$, where $\theta$ and $\phi$ are the measurement angles of the particles.\\In defining these functions one has to notice, that for all possible events (single events, two-particle-events,\ldots) an apropriate function definition has to exist, each taking a list as parameter (e.g. \mmexpr{Prob[\{x\_,y\_\}]=$\frac{1}{2}$Sin[(x\,-\,y)/2]$^2$}).
 \item[\param{inequ-nr (List):}]Range of rows in H-representation used for checking violated inequalities. Specifying this can be useful, if many inequalities have to be evaluated. The form of the parameter is \mmexpr{\{min,max\}} oder \mmexpr{All}.
 \item[\param{violation (Real):}]Only inequalities are printed out, that are violated more than this parameter. Default value is 0.
  \item[\oparam{options:}]Options for the \mathematica-function \cmd{Plot}{\ldots} can be handed over.
  \end{list}
\beschr{Like the function \cmd{PlotInequalities}{\ldots} \cmd{ContPlotInequalities}{\ldots} yields a graphical representation of the violation of Boole-Bell type inequalities, but in this case the functions are dependant on two variables: $f(x,y)=p(x,y)-b$ derived from $p(x,y) \leq b$, where $p(x,y)$ is a linear combination of functions to calculate the probability for single or joint events. Like in the description of the \cmd{PlotInequalities}{\ldots}-function we take the ``2 particles - 2 measurement directions'', the only difference is the selection of the measurement angles: $$a_1 = x\qquad a_2 = b_1 = 0\qquad b_2 = y$$Thus the function $f(x,y)$ is now given by
  \begin{displaymath}
    f(x,y)=1-\frac{1}{2}\sin{\frac{x}{2}}²+\frac{1}{2}\sin{\frac{x-y}{2}}²-\frac{1}{2}\sin{\frac{y}{2}}² - 1
  \end{displaymath}
and can be plotted as contour plots. A higher level of violation is represented by a darker contour layer.}
\beispiel{ContPlotInequalities}{\cmd{ConvToHRep}{2,2},\{x,0,$\pi$\},\{y,0,$\pi$\},\{\{x,0\},\{0,y\}\},Prob,All}{\
  \begin{figure}[H]
    \begin{center}
      \includegraphics[width=50mm]{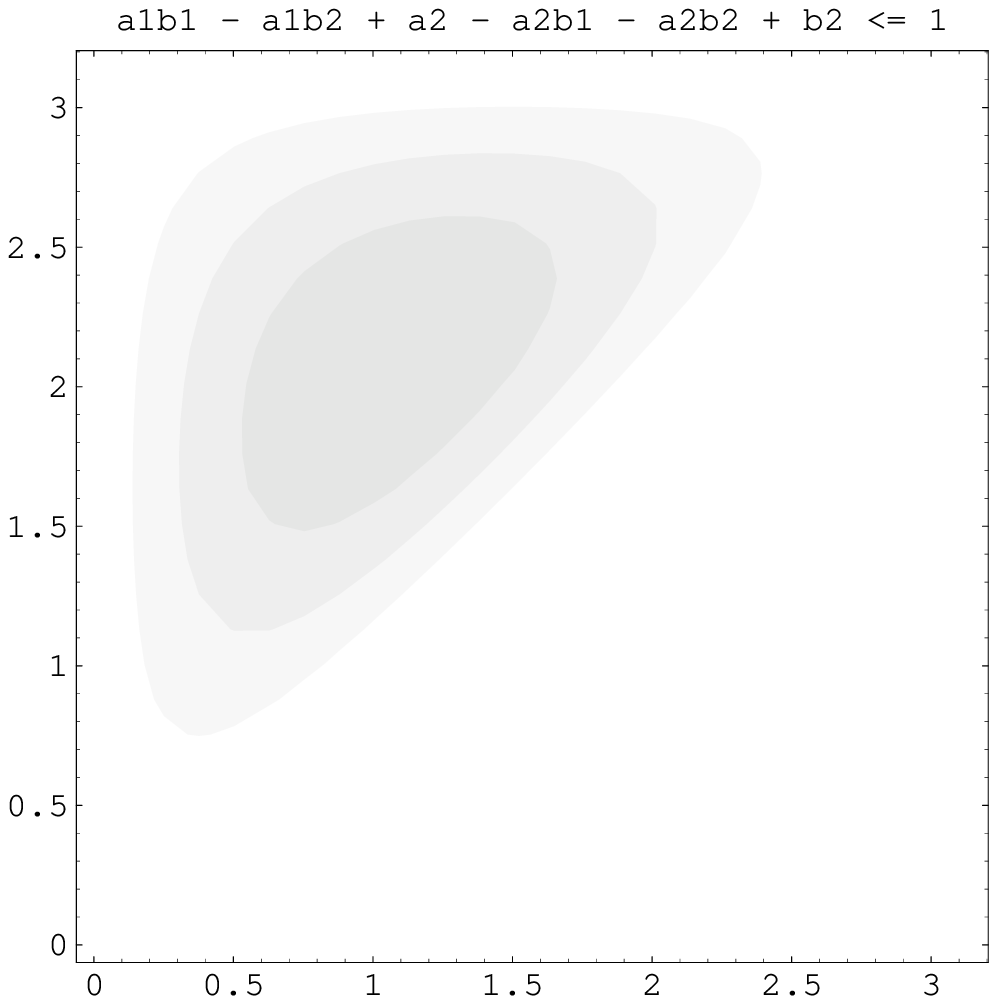}
      \includegraphics[width=50mm]{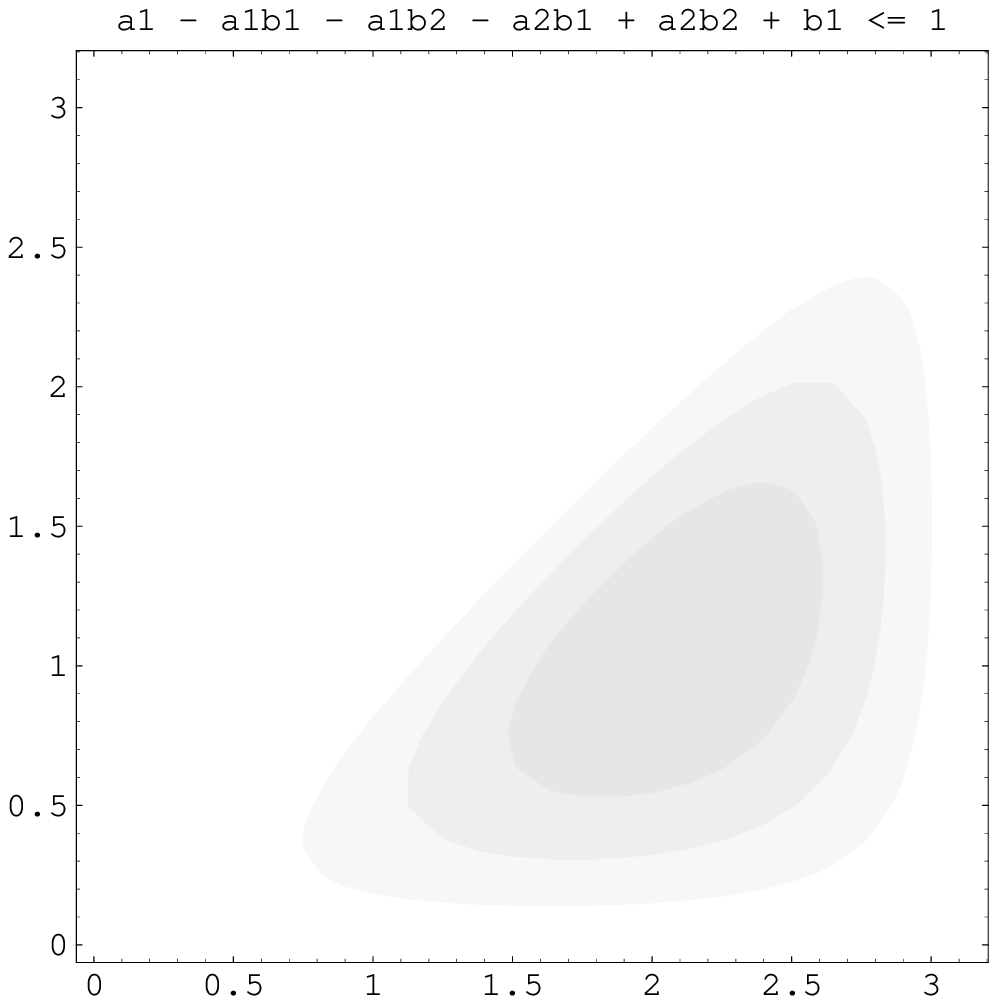}
      \label{fig:cont1}
    \end{center}
  \end{figure}
\par\noindent The functions to calculate the probabilities have been defined as \mmexpr{Prob[\{x\_\}]:=$\frac{1}{2}$} for a single event and \mmexpr{Prob[\{x\_,y\_\}]:=$\frac{1}{2}$Sin[(x\,-\,y)/2]$^2$} for two-particle events.}

\subsection{Cdd}
\label{sec:cdd}

\cddcmd{\cmd{Cdd}{\param{values},\oparam{command}}}
  \begin{list}{param}{}
  \item[\param{values (List):}]Data handed over to \cdd\ in an input-file. Each element of the list must be a string and represents a row in the output-file.
\item[\oparam{command (String):}]Command to be executed on the generated output-file (default is ``cdd'' - the default value can be changed utilizing the function \cmd{SetCddCmd}{\param{newcommand\_String}}). Here one can specify for example ``cddf'' or ``cddr''.
  \end{list}
\beschr{Simple Interface to run \cdd\ in \mathematica. The current version cannot distinguish, whether \cdd\ has produced correct output or not, so one has to pay attention while using this function.}
\beispiel{Cdd}{\param{\{``H-Representation'',``begin'',``6 4 real'',``2 -1 0 0'',``2 0 -1 0'',``-1 1 0 0'',``-1 0 1 0'',``-1 0 0 1'',``4 -1 -1 0'',``end''\}}}
{
  \begin{flushleft}
    \{\{``*'', ``cdd:'', ``Double'', ``Description'', ``Method'', ``C-Code:Version'', ``0.61b'',
    ``(November'', 29, ``1997)''\}, \{``*'', ``Copyright'', ``(C)'', 1996, ``Komei'',
    ``Fukuda,'', ``fukuda@ifor.math.ethz.ch''\}, \{``*Input'', ``File:tmp.ine'', ``('', 6,
     ``x'', ``4)''\}, \{``*HyperplaneOrder:'', ``LexMin''\}, \{``*Degeneracy'',
    ``preknowledge'', ``for'', ``computation:'', ``None'', ``(possible'',
    ``degeneracy)''\}, \{``*Vertex/Ray'', ``enumeration'', ``is'',
    ``chosen.''\}, \{``*Computation'', ``completed'', ``at'', ``Iteration'',
    6.\}, \{``*Computation'', ``starts'', ``at'', ``Thu'', ``Mar'', 22, ``18:48:36'',
    2001\}, \{``*'', ``terminates'', ``at'', ``Thu'', ``Mar'', 22, ``18:48:36'',
    2001\}, \{``*Total'', ``processor'', ``time'', ``='', 0, ``seconds''\}, \{``*'', ``='', 0,
    ``hour'', 0, ``min'', 0, ``sec''\}, \{``*FINAL'', ``RESULT:''\}, \{``*Number'', ``of'',
    ``Vertices'', ``='', 4, ``Rays'', ``='', 1\}, \{``V-representation''\}, \{``begin''\}, \{5,
    4, ``real''\}, \{1, 2, 1, 1\}, \{1, 1, 1, 1\}, \{1, 1, 2, 1\}, \{1, 2, 2, 1\}, \{0, 0,
    0, 1\}, \{``end''\}, \{``hull''\}\}
  \end{flushleft}
}

\section{Examples}
\label{sec:bsp}
These following two examples are originally solved in a paper by Pitowsky and Svozil \cite{2000-poly}.
The associated \mathematica\ - notebooks are ``3\_2.nb''
(three particles - 2 measurement directions) and ``2\_3.nb''
(two particles and three measurement directions)
\subsection{Three particles and two measurement directions}
\label{sec:3_2}
\begin{figure}[!htbp]
  \begin{center}
    \includegraphics[width=80mm]{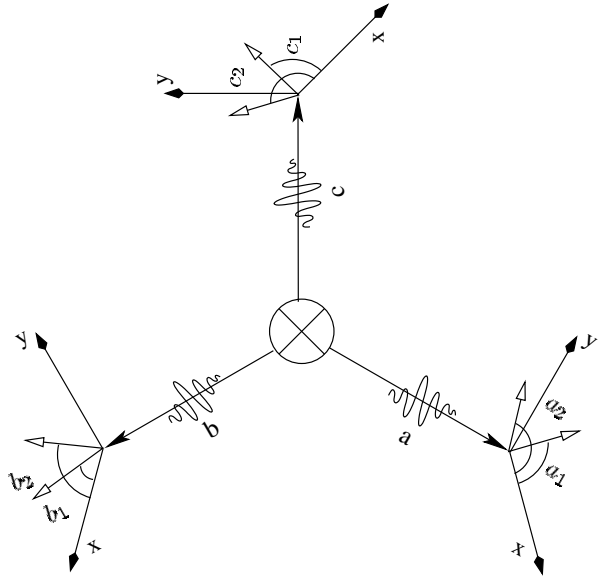}
    \caption{Setting for "2 particles - 3 angles"}
    \label{fig:32graph}
  \end{center}
\end{figure}

In this configuration three particles (\textit{a}, \textit{b} and \textit{c}) are measured in detectors which can be switched between two angles each. Consequently there are six different propositions for single particle events: $a_1$, $a_2$, $b_1$, $b_2$, $c_1$, $c_2$, supposing that $a_1$ is the detection (i. e. the click in a counter) of the particle \textit{a} in the detector set along the axis specified by the first angle for particle \textit{a}, $b_2$ the detection of particle \textit{b} at the second angle for this particle, and so on \ldots (cf. Figure \ref{fig:32graph}).
 If we also take two and three particle events into account (for example the event $a_2c_1$ means a click in the counter for particle \textit{a} at the second angle AND a click in the counter for particle \textit{c} at the first angle), there are 26 different events:
\begin{flushleft}
$a_1$, $a_2$, $b_1$, $b_2$, $c_1$, $c_2$, $a_1b_1$, $a_1b_2$, $a_2b_1$, $a_2b_2$, $a_1c_1$, $a_1c_2$, $a_2c_1$, $a_2c_2$, $b_1c_1$, $b_1c_2$, $b_2c_1$, $b_2$, $c_2$,$a_1b_1c_1$, $a_1b_1c_2$, $a_1b_2c_1$, $a_1b_2c_2$, $a_2b_1c_1$, $a_2b_1c_2$, $a_2b_2c_1$, $a_2b_2c_2$
\end{flushleft}

   \subsection{Violations of inequalities}

The truth table for this configuration can be obtained utilizing the function \cmd{TruthTable}{\param{3,2}} \footnote{due to lack of space not listed here, but it can be found in the \mathematica - notebook \textit{3\_2.nb}}, executing \cmd{ConvToHRep}{3,2} yielded the appropriate H-representation, but this would last quite long, due to the complexity of the correlation polytope for this setting (there are 53856 hyper-planes limiting the polytope). Because of this fact trying to read in the H-representation created by \cdd\ (using \cmd{ReadInHRep}{\ldots}) could also result in memory resource problems by \mathematica.
\par\noindent
To avoid this symptoms it is suggested to export the list of vertices and apply \cdd\ outside of \mathematica\ to the file containing the list of vertices (V-representation). This can be done by invoking \cmd{ToCddExtFile}{\param{``3\_2'',3,2}}, which creates a file ``3\_2.ext''. This file can be handed over to \cdd\ as parameter to get the file ``3\_2.ine'' comprising the H-representation of the correlation polytope (Command: \textit{``cdd 3\_2.ext''}).
\par\noindent
Now the search for violated inequalities can begin using the function \cmd{GetViolInequalities}{\param{\ldots}}:
\par\noindent
May be accepted that the functions to calculate the quantum probabilities of the (joint) events (\mmexpr{Prob}) have been defined by \mmexpr{Prob[\{x\_\}]:=$\frac{1}{2}$},\ \mmexpr{Prob[\{x\_,y\_\}]:=$\frac{1}{4}$} and \mmexpr{Prob[\{x\_,y\_,z\_\}]:=$\frac{1}{8}*(1-Sin[x+y+z])$}, where x, y and z are the angles used for detection of each particle,
\begin{center}
\cmd{GetViolInequalities}{``3\_2.ine'',\{\{0,$\frac{\pi}{2}$\},\{0,$\frac{\pi}{2}$\},\{0,$\frac{\pi}{2}$\}\},Prob,All,0.4}
\end{center}
yields:

\begin{longtable}[t]{l}

\{\{-3\  \mmexpr{a1}+2\  \mmexpr{a1b1}+\mmexpr{a1b1c1}-4\  \mmexpr{a1b1c2}+3\  \mmexpr{a1b2}-3\  \mmexpr{a1b2c1}-
 \\
\noalign{\vspace{0.571429ex}}
\hspace{4.em} \mmexpr{a1b2c2}+\mmexpr{a1c1}+3\  \mmexpr{a1c2}+2\  \mmexpr{a2b1}-2\  \mmexpr{a2b1c1}-\mmexpr{a2b1c2}-
 \\
\noalign{\vspace{0.571429ex}}
\hspace{4.em} 2\  \mmexpr{a2b2}+\mmexpr{a2b2c1}+3\  \mmexpr{a2b2c2}+\mmexpr{a2c1}-\mmexpr{a2c2}-2\  \mmexpr{b1}+
 \\
\noalign{\vspace{0.571429ex}}
\hspace{4.em} \mmexpr{b1c1}+2\  \mmexpr{b1c2}+\mmexpr{b2c1}-2\  \mmexpr{b2c2}-\mmexpr{c1} $\leq$ 0,0.5\},  \\
\noalign{\vspace{0.571429ex}}
\hspace{1.em} \{-2\  \mmexpr{a1}+2\  \mmexpr{a1b1}+\mmexpr{a1b1c1}-4\  \mmexpr{a1b1c2}+2\  \mmexpr{a1b2}-2\
\mmexpr{a1b2c1}-  \\
\noalign{\vspace{0.571429ex}}
\hspace{4.em} \mmexpr{a1b2c2}+\mmexpr{a1c1}+2\  \mmexpr{a1c2}+3\  \mmexpr{a2b1}-3\  \mmexpr{a2b1c1}-\mmexpr{a2b1c2}-
 \\
\noalign{\vspace{0.571429ex}}
\hspace{4.em} 2\  \mmexpr{a2b2}+\mmexpr{a2b2c1}+3\  \mmexpr{a2b2c2}+\mmexpr{a2c1}-2\  \mmexpr{a2c2}-3\  \mmexpr{b1}+
 \\
\noalign{\vspace{0.571429ex}}
\hspace{4.em} \mmexpr{b1c1}+3\  \mmexpr{b1c2}+\mmexpr{b2c1}-\mmexpr{b2c2}-\mmexpr{c1} $\leq$ 0,0.5\},\newline   \\
\noalign{\vspace{0.571429ex}}
\hspace{1.em} \{-2\  \mmexpr{a1}+\mmexpr{a1b1}+\mmexpr{a1b1c1}-4\  \mmexpr{a1b1c2}+2\  \mmexpr{a1b2}-3\  \mmexpr{a1b2c1}-
 \\
\noalign{\vspace{0.571429ex}}
\hspace{4.em} \mmexpr{a1b2c2}+2\  \mmexpr{a1c1}+2\  \mmexpr{a1c2}+2\  \mmexpr{a2b1}-3\  \mmexpr{a2b1c1}-  \\
\noalign{\vspace{0.571429ex}}
\hspace{4.em} \mmexpr{a2b1c2}-\mmexpr{a2b2}+\mmexpr{a2b2c1}+2\  \mmexpr{a2b2c2}+\mmexpr{a2c1}-\mmexpr{a2c2}-  \\
\noalign{\vspace{0.571429ex}}
\hspace{4.em} 2\  \mmexpr{b1}+2\  \mmexpr{b1c1}+2\  \mmexpr{b1c2}+\mmexpr{b2c1}-\mmexpr{b2c2}-2\  \mmexpr{c1} $\leq$ 0,0.5\},
 \\
\noalign{\vspace{0.571429ex}}
\hspace{1.em} \{-2\  \mmexpr{a1}+2\  \mmexpr{a1b1}+\mmexpr{a1b1c1}-4\  \mmexpr{a1b1c2}+2\  \mmexpr{a1b2}-3\
\mmexpr{a1b2c1}-  \\
\noalign{\vspace{0.571429ex}}
\hspace{4.em} \mmexpr{a1b2c2}+3\  \mmexpr{a1c2}+2\  \mmexpr{a2b1}-3\  \mmexpr{a2b1c1}-\mmexpr{a2b1c2}-  \\
\noalign{\vspace{0.571429ex}}
\hspace{4.em} 2\  \mmexpr{a2b2}+\mmexpr{a2b2c1}+2\  \mmexpr{a2b2c2}+\mmexpr{a2c1}-2\  \mmexpr{b1}+2\  \mmexpr{b1c1}+
 \\
\noalign{\vspace{0.571429ex}}
\hspace{4.em} 2\  \mmexpr{b1c2}+\mmexpr{b2c1}-\mmexpr{b2c2}-\mmexpr{c1}-\mmexpr{c2} $\leq$ 0,0.5\},
\\
\noalign{\vspace{0.571429ex}}
\hspace{1.em} \{-\mmexpr{a1}+\mmexpr{a1b1}+\mmexpr{a1b1c1}-4\  \mmexpr{a1b1c2}+\mmexpr{a1b2}-3\  \mmexpr{a1b2c1}-  \\
\noalign{\vspace{0.571429ex}}
\hspace{4.em} \mmexpr{a1b2c2}+\mmexpr{a1c1}+2\  \mmexpr{a1c2}+\mmexpr{a2}+\mmexpr{a2b1}-2\  \mmexpr{a2b1c1}-  \\
\noalign{\vspace{0.571429ex}}
\hspace{4.em} \mmexpr{a2b1c2}-2\  \mmexpr{a2b2}+\mmexpr{a2b2c1}+3\  \mmexpr{a2b2c2}-\mmexpr{a2c2}-\mmexpr{b1}+  \\
\noalign{\vspace{0.571429ex}}
\hspace{4.em} \mmexpr{b1c1}+2\  \mmexpr{b1c2}+\mmexpr{b2}+\mmexpr{b2c1}-2\  \mmexpr{b2c2}-\mmexpr{c1} $\leq$ 1,1.5\},
\\
\noalign{\vspace{0.571429ex}}
\hspace{1.em} \{-\mmexpr{a1}+2\  \mmexpr{a1b1}-\mmexpr{a1b1c1}-3\  \mmexpr{a1b1c2}+\mmexpr{a1b2}-4\  \mmexpr{a1b2c1}+
 \\
\noalign{\vspace{0.571429ex}}
\hspace{4.em} \mmexpr{a1b2c2}+3\  \mmexpr{a1c1}+\mmexpr{a2}+\mmexpr{a2b1}-2\  \mmexpr{a2b1c1}-\mmexpr{a2b1c2}-  \\
\noalign{\vspace{0.571429ex}}
\hspace{4.em} 2\  \mmexpr{a2b2}+\mmexpr{a2b2c1}+3\  \mmexpr{a2b2c2}-\mmexpr{a2c2}-2\  \mmexpr{b1}+2\  \mmexpr{b1c1}+
 \\
\noalign{\vspace{0.571429ex}}
\hspace{4.em} 2\  \mmexpr{b1c2}+\mmexpr{b2}+2\  \mmexpr{b2c1}-3\  \mmexpr{b2c2}-2\  \mmexpr{c1}+\mmexpr{c2} $\leq$ 1,1.5\},
 \\
\noalign{\vspace{0.571429ex}}
\hspace{1.em} \{-2\  \mmexpr{a1}+2\  \mmexpr{a1b1}-\mmexpr{a1b1c1}-3\  \mmexpr{a1b1c2}+\mmexpr{a1b2}-2\  \mmexpr{a1b2c1}-
 \\
\noalign{\vspace{0.571429ex}}
\hspace{5.em} \mmexpr{a1b2c2}+2\  \mmexpr{a1c1}+2\  \mmexpr{a1c2}+\mmexpr{a2}+\mmexpr{a2b1}-4\  \mmexpr{a2b1c1}+
 \\
\noalign{\vspace{0.571429ex}}
\hspace{5.em} \mmexpr{a2b1c2}-2\  \mmexpr{a2b2}+\mmexpr{a2b2c1}+3\  \mmexpr{a2b2c2}+2\  \mmexpr{a2c1}-3\  \mmexpr{a2c2}-
 \\
\noalign{\vspace{0.571429ex}}
\hspace{5.em} \mmexpr{b1}+3\  \mmexpr{b1c1}+\mmexpr{b2}-\mmexpr{b2c2}-2\  \mmexpr{c1}+\mmexpr{c2} $\leq$ 1,1.5\}
\\
\noalign{\vspace{0.571429ex}}
\hspace{1.em} \{\ldots\}\ldots\}

\end{longtable}
\addtocounter{table}{-1}

   \subsection{Graphical representation}

Using \cmd{PlotInequalities}{\ldots} a graph can be created showing the violation of inequalities dependent on one variable. Defining the probability functions as above, executing
\begin{center}
  \cmd{PlotInequalities}{\param{``3\_2.ine'',\{x,0,$\pi$\},\{\{0,x\},\{0,x\},\{0,x\},Prob,\{10000,20000\},0.4}}
\end{center}
yields the following plot (cf. Figure \ref{fig:32plot}), whereas the corresponding H-representation has to be stored in the file \textit{``3\_2.ine''}, \{10000,20000\} indicates the range of row numbers taken for calculating the graph and \mmexpr{0.4} is the minimal degree of violation to include the inequality in the graph:
\begin{figure}[!htbp]
  \begin{center}
    \includegraphics[width=70mm]{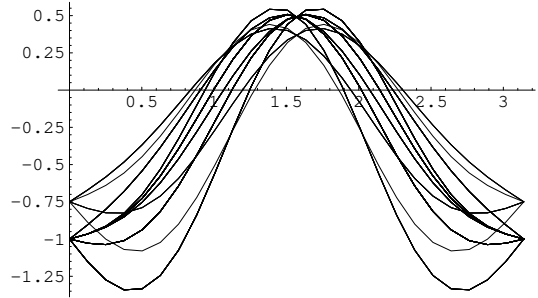}
    \caption{PlotInequalities[``3\_2.ine'',\{x,0,$\pi$\},\{\{0,x\},\{0,x\},\{0,x\},Prob,\{10000,20000\},0.4]}
    \label{fig:32plot}
  \end{center}
\end{figure}
\par
\begin{figure}[!htbp]
  \begin{center}
    \includegraphics[width=50mm]{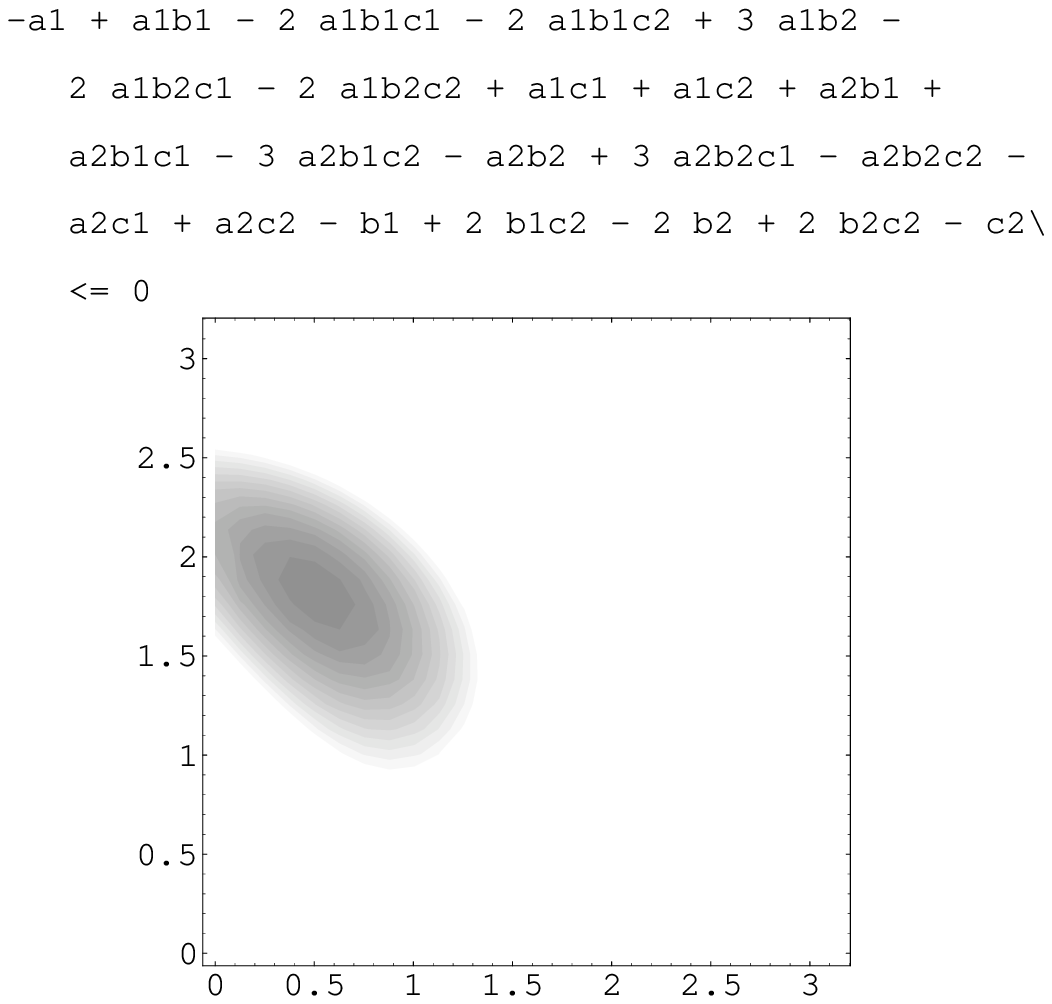}
    \includegraphics[width=50mm]{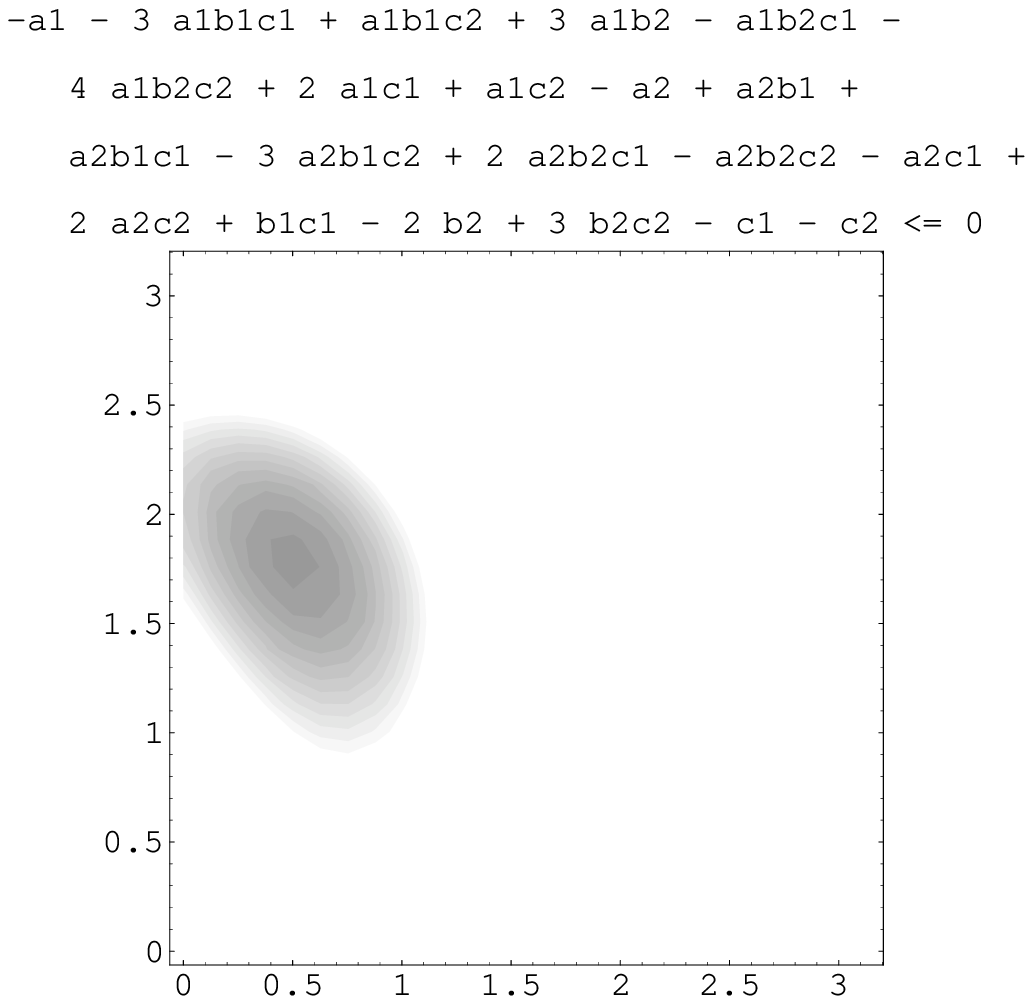}\\
    \vspace{1cm}
    \includegraphics[width=50mm]{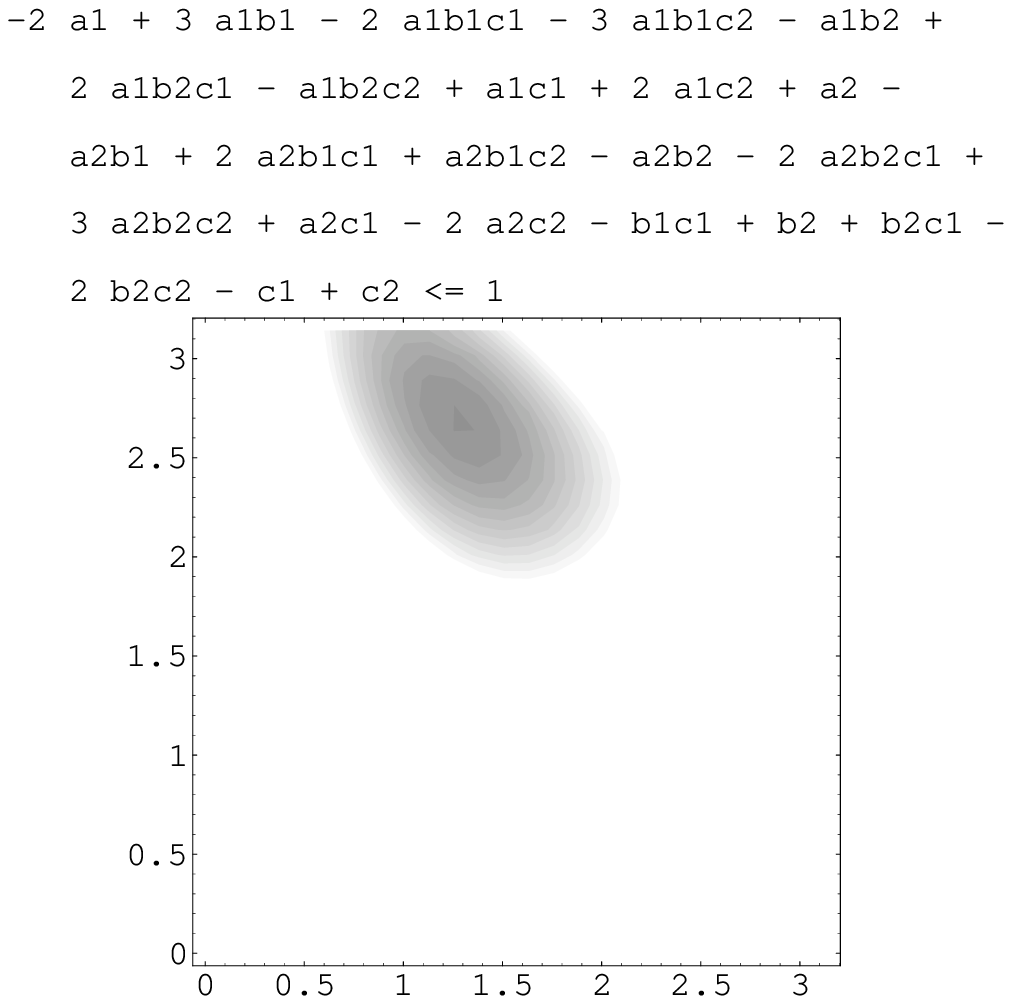}
    \includegraphics[width=50mm]{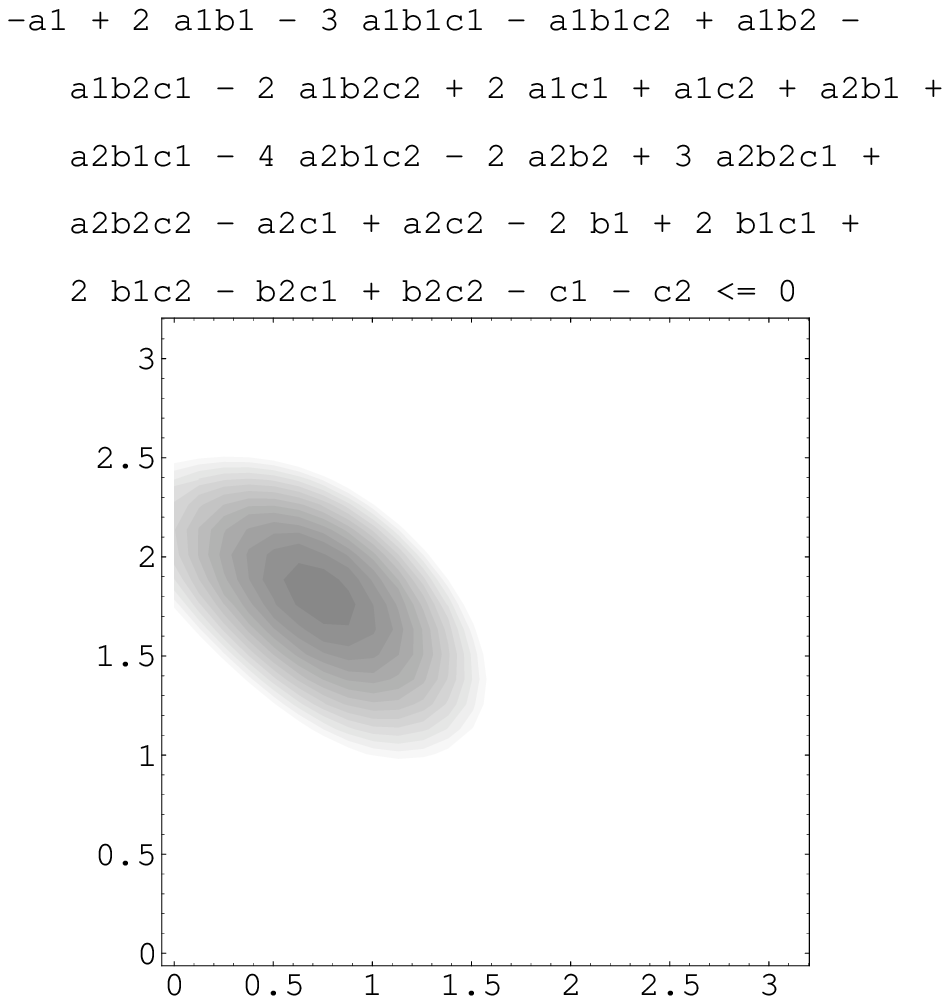}
    \caption{ContPlotInequalities[``3\_2.ine'',\{x,0,$\pi$\},\{x,0,$\pi$\},\{\{0,x\},\{0,y\},\{x,y\}\},Prob,\{10000,20000\},0.4]\newline Show[GraphicsArray[Partition[cont,3][[\{1,2\},All]] // TableForm}
    \label{fig:32cont}
  \end{center}
\end{figure}

\noindent
To display inequalities dependent on two variables the function \cmd{ContPlotInequalities}{\ldots} is provided. This function shows the violation as a contour plot, a more violated set of detection angles results in a darker region in the plot.
\begin{center}
  \cmd{ContPlotInequalities}{\param{``3\_2.ine'',\{x,0,$\pi$\},\{x,0,$\pi$\},\{\{0,x\},\{0,y\},\{x,y\}\},Prob,\{10000,20000\},0.4}}
\end{center}
returns \mmexpr{ContourGraphics}-objects, which can be displayed for example by executing
\begin{center}
  \cmd{Show}{\cmd{GraphicsArray}{\cmd{Partition}{\param{cont, 3}\mmexpr{[[\param{\{1,2\},All}]]}}}}\mmexpr{\ //\ TableForm}
\end{center}
(see figure \ref{fig:32cont}).

\subsection{Two particles and three measurement directions}
\label{sec:3_2a}
\begin{figure}[!htbp]
  \begin{center}
    \includegraphics[width=80mm]{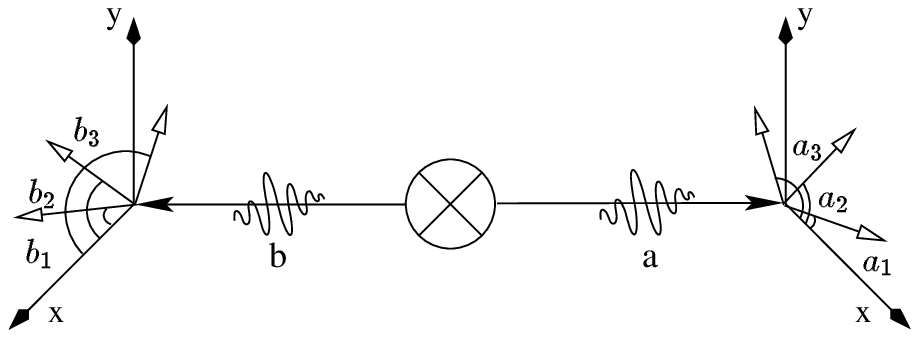}
    \caption{Setting for "2 particles - 3 angles"}
    \label{fig:23graph}
  \end{center}
\end{figure}

In the case of two particles (\textit{a} and \textit{b}) with three properties (whereas the properties are three different angles of the detectors for each particle denoted by $a_1$, $a_2$, $a_3$, $b_1$, $b_2$, $b_3$ - see figure \ref{fig:23graph}) 15 different events can be found:
\begin{flushleft}
\{$a_1$, $a_2$, $a_3$, $b_1$, $b_2$, $b_3$, $c_1$, $c_2$, $c_3$, $a_1b_1$, $a_1b_2$, $a_1b_3$, $a_2b_1$, $a_2b_2$, $a_2b_3$,  $a_3b_1$, $a_3b_2$, $a_3b_3$\}
\end{flushleft}
   \subsection{Violations of inequalities}
Using \cmd{TruthTable}{2,3} all vertices of the corresponding correlation polytope can be found -  we get a dimension of 15 and 64 vertices as result (table \ref{tab:tt6}).

\begin{scriptsize}
\renewcommand{\baselinestretch}{0.7}
\begin{longtable}[t]{*{15}{c}}
  \mmexpr{a1}&\mmexpr{a2}&\mmexpr{a3}&\mmexpr{b1}&\mmexpr{b2}&\mmexpr{b3}&\mmexpr{a1b1}&\mmexpr{a1b2}&\mmexpr{a1b3}&\mmexpr{a2b1}&\mmexpr{a2b2}&\mmexpr{a2b3}&\mmexpr{a3b1}&\mmexpr{a3b2}&\mmexpr{a3b3}\\
  \hline
  \endhead
  0&0&0&0&0&0&0&0&0&0&0&0&0&0&0 \\
  1&0&0&0&0&0&0&0&0&0&0&0&0&0&0 \\
  0&1&0&0&0&0&0&0&0&0&0&0&0&0&0 \\
  1&1&0&0&0&0&0&0&0&0&0&0&0&0&0 \\
  0&0&1&0&0&0&0&0&0&0&0&0&0&0&0 \\
  1&0&1&0&0&0&0&0&0&0&0&0&0&0&0 \\
  0&1&1&0&0&0&0&0&0&0&0&0&0&0&0 \\
  1&1&1&0&0&0&0&0&0&0&0&0&0&0&0 \\
  0&0&0&1&0&0&0&0&0&0&0&0&0&0&0 \\
  1&0&0&1&0&0&1&0&0&0&0&0&0&0&0 \\
  0&1&0&1&0&0&0&0&0&1&0&0&0&0&0 \\
  1&1&0&1&0&0&1&0&0&1&0&0&0&0&0 \\
  0&0&1&1&0&0&0&0&0&0&0&0&1&0&0 \\
  1&0&1&1&0&0&1&0&0&0&0&0&1&0&0 \\
  0&1&1&1&0&0&0&0&0&1&0&0&1&0&0 \\
  1&1&1&1&0&0&1&0&0&1&0&0&1&0&0 \\
  0&0&0&0&1&0&0&0&0&0&0&0&0&0&0 \\
  1&0&0&0&1&0&0&1&0&0&0&0&0&0&0 \\
  0&1&0&0&1&0&0&0&0&0&1&0&0&0&0 \\
  1&1&0&0&1&0&0&1&0&0&1&0&0&0&0 \\
  0&0&1&0&1&0&0&0&0&0&0&0&0&1&0 \\
  1&0&1&0&1&0&0&1&0&0&0&0&0&1&0 \\
  0&1&1&0&1&0&0&0&0&0&1&0&0&1&0 \\
  1&1&1&0&1&0&0&1&0&0&1&0&0&1&0 \\
  0&0&0&1&1&0&0&0&0&0&0&0&0&0&0 \\
  1&0&0&1&1&0&1&1&0&0&0&0&0&0&0 \\
  0&1&0&1&1&0&0&0&0&1&1&0&0&0&0 \\
  1&1&0&1&1&0&1&1&0&1&1&0&0&0&0 \\
  0&0&1&1&1&0&0&0&0&0&0&0&1&1&0 \\
  1&0&1&1&1&0&1&1&0&0&0&0&1&1&0 \\
  0&1&1&1&1&0&0&0&0&1&1&0&1&1&0 \\
  1&1&1&1&1&0&1&1&0&1&1&0&1&1&0 \\
  0&0&0&0&0&1&0&0&0&0&0&0&0&0&0 \\
  1&0&0&0&0&1&0&0&1&0&0&0&0&0&0 \\
  0&1&0&0&0&1&0&0&0&0&0&1&0&0&0 \\
  1&1&0&0&0&1&0&0&1&0&0&1&0&0&0 \\
  0&0&1&0&0&1&0&0&0&0&0&0&0&0&1 \\
  1&0&1&0&0&1&0&0&1&0&0&0&0&0&1 \\
  0&1&1&0&0&1&0&0&0&0&0&1&0&0&1 \\
  1&1&1&0&0&1&0&0&1&0&0&1&0&0&1 \\
  0&0&0&1&0&1&0&0&0&0&0&0&0&0&0 \\
  1&0&0&1&0&1&1&0&1&0&0&0&0&0&0 \\
  0&1&0&1&0&1&0&0&0&1&0&1&0&0&0 \\
  1&1&0&1&0&1&1&0&1&1&0&1&0&0&0 \\
  0&0&1&1&0&1&0&0&0&0&0&0&1&0&1 \\
  1&0&1&1&0&1&1&0&1&0&0&0&1&0&1 \\
  0&1&1&1&0&1&0&0&0&1&0&1&1&0&1 \\
  1&1&1&1&0&1&1&0&1&1&0&1&1&0&1 \\
  0&0&0&0&1&1&0&0&0&0&0&0&0&0&0 \\
  1&0&0&0&1&1&0&1&1&0&0&0&0&0&0 \\
  0&1&0&0&1&1&0&0&0&0&1&1&0&0&0 \\
  1&1&0&0&1&1&0&1&1&0&1&1&0&0&0 \\
  0&0&1&0&1&1&0&0&0&0&0&0&0&1&1 \\
  1&0&1&0&1&1&0&1&1&0&0&0&0&1&1 \\
  0&1&1&0&1&1&0&0&0&0&1&1&0&1&1 \\
  1&1&1&0&1&1&0&1&1&0&1&1&0&1&1 \\
  0&0&0&1&1&1&0&0&0&0&0&0&0&0&0 \\
  1&0&0&1&1&1&1&1&1&0&0&0&0&0&0 \\
  0&1&0&1&1&1&0&0&0&1&1&1&0&0&0 \\
  1&1&0&1&1&1&1&1&1&1&1&1&0&0&0 \\
  0&0&1&1&1&1&0&0&0&0&0&0&1&1&1 \\
  1&0&1&1&1&1&1&1&1&0&0&0&1&1&1 \\
  0&1&1&1&1&1&0&0&0&1&1&1&1&1&1 \\
  1&1&1&1&1&1&1&1&1&1&1&1&1&1&1 \\
  \caption{\protect\rule{0cm}{20pt}Truth table for 6 propositions}
  \label{tab:tt6}
\end{longtable}
\end{scriptsize}
\par
Executing \cmd{hrep=ConvToHRep}{2,3} a H-representation of the polytope described by the truth table above can be created. This results in 684 hyper-planes respectively 684 inequalities from the 64 vertices limiting the polytope:
\begin{longtable}[t]{l}
  \mmexpr{\{\{H-representation\}, \{begin\},\{684, 16, real\},}\\
  \noalign{\vspace{0.571429ex}}
  \mmexpr{\{2, 0, -2, 1, -1, 0, -1, 1, -1, 0, 1, 1, 1, -1, -1, 1\},}  \\
  \noalign{\vspace{0.571429ex}}
  \mmexpr{\{2, -2, 0, 1, -1, 0, -1, 1, 1, 1, 1, -1, 0, -1, -1, 1\},} \\
  \noalign{\vspace{0.571429ex}}
  \mmexpr{\{\ldots\},} \\
  \noalign{\vspace{0.571429ex}}
  \mmexpr{\{1, -1, 0, 0, -1, 0, 0, 1, 0, 0, 0, 0, 0, 0, 0, 0\},} \\
  \noalign{\vspace{0.571429ex}}
  \mmexpr{\{end\},\{Konfiguration,2,3\}\}}
\end{longtable}
\noindent All inequalities can be displayed by
\begin{center}
\cmd{GetInequFromHRep}{hrep}\mmexpr{// InequToRead}
\end{center}
The result of this operation is
\begin{scriptsize}
\begin{longtable}[t]{l}
  \mmexpr{-a1b1 + a1b2 + 2\ a2 - a2b1 - a2b2 - a2b3 - a3 + a3b1 + a3b2 - a3b3 + b1 + b3} $\leq$ \mmexpr{2}\\
  \noalign{\vspace{0.571429ex}}
  \mmexpr{2a1 - a1b1 - a1b2 - a1b3 - a2b1 + a2b2 - a3 + a3b1 + a3b2 - a3b3 + b1 + b3} $\leq$ \mmexpr{2}\\
  \noalign{\vspace{0.571429ex}}
  \mmexpr{a1 - a1b1 - a1b3 - a2b1 + a2b2 + a2b3 - a3 + a3b1 + a3b2 - a3b3 + b1 - b2} $\leq$ \mmexpr{1}\\
  \noalign{\vspace{0.571429ex}}
  \mmexpr{\ldots}\\
  \noalign{\vspace{0.571429ex}}
  \mmexpr{\ldots}\\
  \noalign{\vspace{0.571429ex}}
  \mmexpr{a1- a1b1 + b1} $\leq$ \mmexpr{1}
\end{longtable}
\noindent
\end{scriptsize}
Using
\begin{center}
\cmd{GetViolInequalities}{hrep,\{\{0,$\frac{2\pi}{3}$,$\frac{4\pi}{3}$\},\{\{0,$\frac{2\pi}{3}$,$\frac{4\pi}{3}$\}\},Prob,All}\mmexpr{// TableForm}
\end{center}
all inequalities can be displayed that are violated at the specific angles
$a_1 = b_1 = 0$, $a_2 = b_2 = \frac{2\pi}{3}$ and $a_3 = b_3 = \frac{4\pi}{3}$
taking the functions \mmexpr{Prob[\{x\_\}]:=$\frac{1}{2}$}
and \mmexpr{Prob[\{x\_\,,y\_\}]:=$Sin[\frac{x-y}{2}]/2$} to calculate the quantum
probabilities, which is equivalent to the probability to find two
spin-$\frac{1}{2}$ particles in a singlet state
($|\psi\rangle = \frac{1}{\sqrt{2}}(|\uparrow\downarrow\rangle - |\downarrow\uparrow\rangle)$)
both either in spin ``up'' or both in spin ``down''.

\begin{scriptsize}
\begin{displaymath}
  \begin{array}[l]{cc}
    \mmexpr{-a1 - a1b1 + a1b2 + a1b3 - a2 + a2b1 + a2b3 + a3b1 + a3b2 - a3b3 - b1 - b2} \leq \mmexpr{0} &\frac{1}{4}\\
    \noalign{\vspace{0.571429ex}}
    \mmexpr{-a1 - a1b1 + a1b2 + a1b3 + a2b1 - a2b2 + a2b3 - a3 + a3b1 + a3b2 - b1 - b3} \leq \mmexpr{0} &\frac{1}{4}\\
    \noalign{\vspace{0.571429ex}}
    \mmexpr{-a1 + a1b2 + a1b3 - a2 + a2b1 - a2b2 + a2b3 + a3b1 + a3b2 - a3b3 - b1 - b2} \leq \mmexpr{0} & \frac{1}{4}\\
    \noalign{\vspace{0.571429ex}}
    \mmexpr{-a1b1 + a1b2 + a1b3 - a2 + a2b1 - a2b2 + a2b3 - a3 + a3b1 + a3b2 - b2 - b3} \leq \mmexpr{0} & \frac{1}{4}\\
    \noalign{\vspace{0.571429ex}}
    \mmexpr{-a1 + a1b2 + a1b3 + a2b1 - a2b2 + a2b3 - a3 + a3b1 + a3b2 - a3b3 - b1 - b3} \leq \mmexpr{0} & \frac{1}{4}\\
    \noalign{\vspace{0.571429ex}}
    \mmexpr{-a1b1 + a1b2 + a1b3 - a2 + a2b1 + a2b3 - a3 + a3b1 + a3b2 - a3b3 - b2 - b3} \leq \mmexpr{0} & \frac{1}{4}\\
    \noalign{\vspace{0.571429ex}}
    \mmexpr{-a1 + a1b2 + a1b3 - a2b2 + a2b3 - b3} \leq \mmexpr{0} &\frac{1}{8}\\
    \noalign{\vspace{0.571429ex}}
    \mmexpr{-a1b1 + a1b3 - a2 + a2b1 + a2b3 - b3} \leq \mmexpr{0} &\frac{1}{8}\\
    \noalign{\vspace{0.571429ex}}
    \mmexpr{-a1b1 + a1b2 - a3 + a3b1 + a3b2 - b2} \leq \mmexpr{0} &\frac{1}{8}\\
    \noalign{\vspace{0.571429ex}}
    \mmexpr{-a1 + a1b2 + a1b3 + a3b2 - a3b3 - b2} \leq \mmexpr{0} &\frac{1}{8}\\
    \noalign{\vspace{0.571429ex}}
    \mmexpr{a2b1 - a2b2 - a3 + a3b1 + a3b2 - b1} \leq \mmexpr{0} &\frac{1}{8}\\
    \noalign{\vspace{0.571429ex}}
    \mmexpr{-a2 + a2b1 + a2b3 + a3b1 - a3b3 - b1} \leq \mmexpr{0} &\frac{1}{8}
  \end{array}
\end{displaymath}
\end{scriptsize}

   \subsection{Graphical representation}
Like in the configuration ``three particles and two angles'' described above a graphical
representation can be generated either dependent on one or dependent on two variables.
\par
\begin{figure}[H]
  \begin{center}
    \includegraphics[width=70mm]{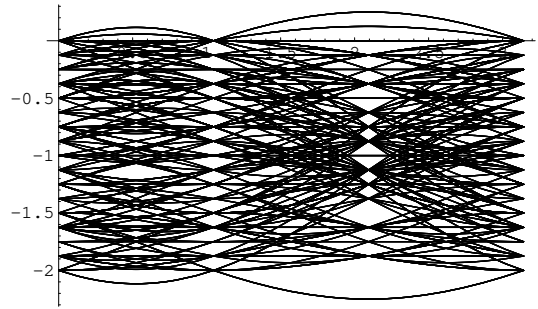}
    \caption{PlotInequalities[hrep,\{x,0,$\pi$\},\{\{0,x,$\frac{4\pi}{3}$\},\{\{0,x,$\frac{4\pi}{3}$\}\},Prob]}
    \label{fig:23plot}
  \end{center}
\end{figure}

\noindent
On the one hand side by using
\begin{center}
\cmd{PlotInequalities}{hrep,\{x,0,$\pi$\},\{\{0,x,$\frac{4\pi}{3}$\},\{\{0,x,$\frac{4\pi}{3}$\}\},Prob}
\end{center}
we get a plot of all violated inequalities (cf. Figure \ref{fig:23plot}),
\noindent on the other hand side contour plots of all inequalities violated for example more than 0.2 can be generated by executing
\begin{center}
\cmd{cont=ContPlotInequalities}{hrep,\{x,0,$\pi$\},\{y,0,$\pi$\},\{\{0,x,y\},\{\{0,x,y\}\},Prob,All,0.2}
\end{center}
To display the outcome of the calculation  entering \mmexpr{Show /@ cont} results in ContourPlots of the following form (cf. Figure \ref{fig:23cont}):
\begin{figure}[H]
  \begin{center}
    \includegraphics[width=45mm]{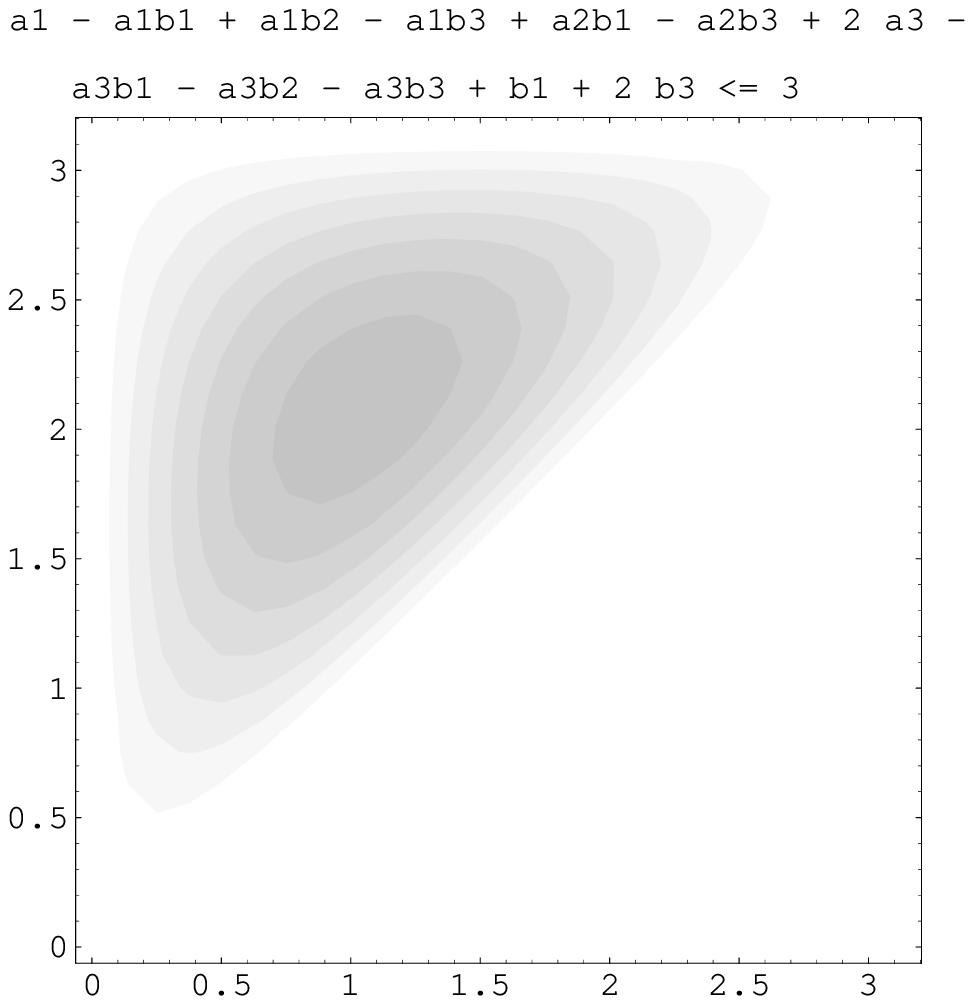}
    \includegraphics[width=45mm]{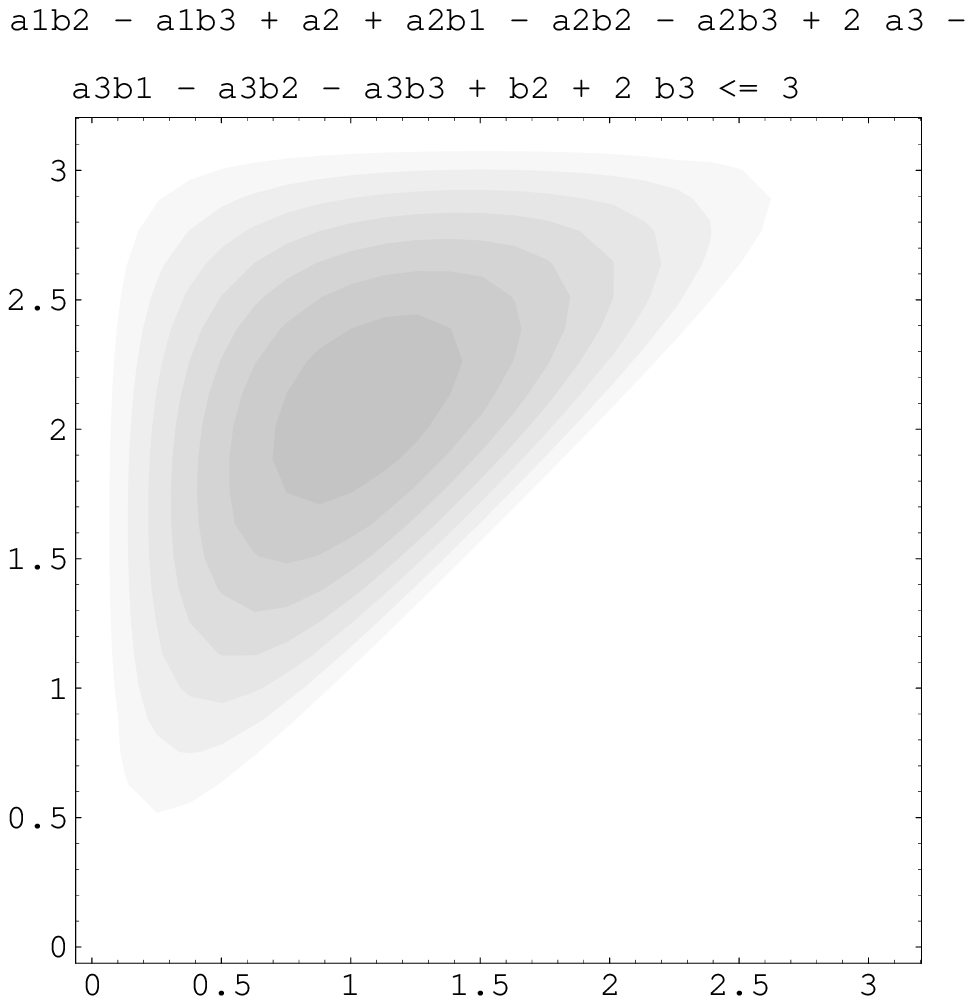}\\
    \vspace{5mm}
    \includegraphics[width=45mm]{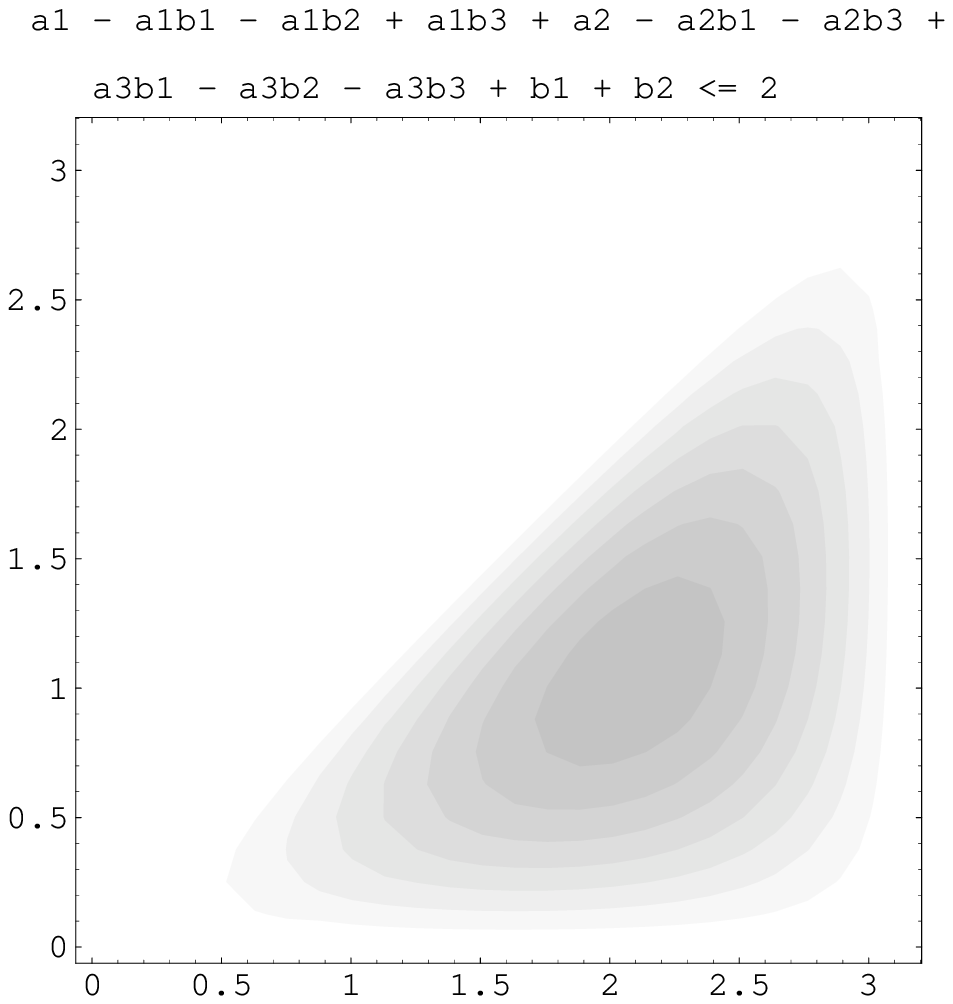}
    \includegraphics[width=45mm]{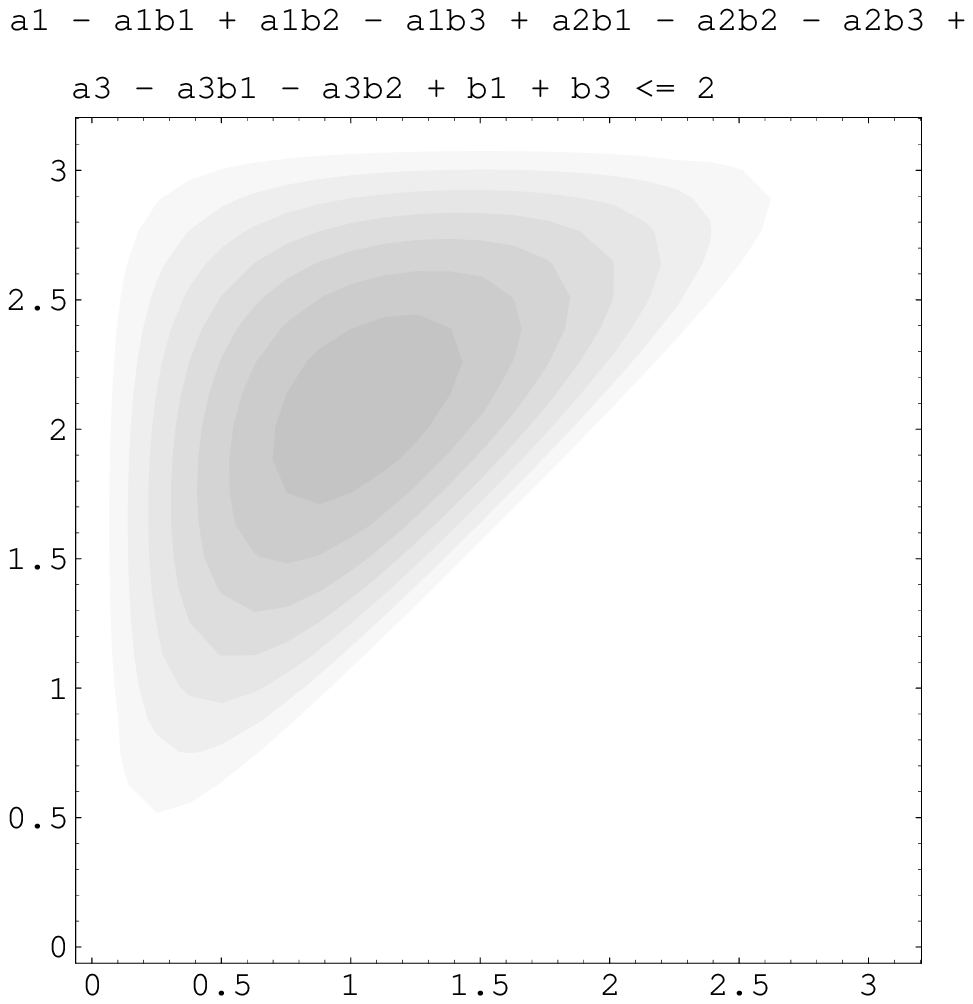}
    \caption{ContPlotInequalities[hrep,\{x,0,$\pi$\},\{y,0,$\pi$\},\{\{0,x,y\},\{\{0,x,y\}\},Prob,All,0.2]\newline Show /@ \%}
    \label{fig:23cont}
  \end{center}
\end{figure}


\begin{thebibliography}{10}

\bibitem{Boole}
George Boole.
\newblock {\em An investigation of the laws of thought}.
\newblock Dover edition, New York, 1958.

\bibitem{Boole-62}
George Boole.
\newblock On the theory of probabilities.
\newblock {\em Philosophical Transactions of the Royal Society of London},
  152:225--252, 1862.

\bibitem{Hailperin}
Theodore Hailperin.
\newblock {\em Boole's logic and probability (Studies in logic and the
  foundations of mathematics ; 85)}.
\newblock North-Holland, Amsterdam, 1976.

\bibitem{pitowsky}
Itamar Pitowsky.
\newblock {\em Quantum Probability---Quantum Logic}.
\newblock Springer, Berlin, 1989.

\bibitem{Pit-94}
Itamar Pitowsky.
\newblock {G}eorge {B}oole's `conditions od possible experience' and the
  quantum puzzle.
\newblock {\em Brit. J. Phil. Sci.}, 45:95--125, 1994.

\bibitem{cl-horne}
J.~F. Clauser and M.~A. Horne.
\newblock {\em Physical Review}, D10:526, 1974.

\bibitem{chsh}
J.~F. Clauser, M.~A. Horne, A.~Shimony, and R.~A. Holt.
\newblock {\em Physical Review Letters}, 23:880--884, 1969.

\bibitem{clauser}
J.~F. Clauser and A.~Shimony.
\newblock {B}ellos theorem: experimental tests and implications.
\newblock {\em Rep. Prog. Phys.}, 41:1881--1926, 1978.

\bibitem{pitowsky-89a}
Itamar Pitowsky.
\newblock From {G}eorge {B}oole to {J}ohn {B}ell: The origin of {B}ell's
  inequality.
\newblock In M.~Kafatos, editor, {\em {B}ell's Theorem, Quantum Theory and the
  Conceptions of the Universe}. Kluwer, Dordrecht, 1989.

\bibitem{Pit-91}
Itamar Pitowsky.
\newblock Correlation polytopes their geometry and complexity.
\newblock {\em Mathematical Programming}, 50:395--414, 1991.

\bibitem{ziegler}
G{\"{u}}nter~M. Ziegler.
\newblock {\em Lectures on Polytopes}.
\newblock Springer, New York, 1994.

\bibitem{MRTT53}
T.S. Motzkin, H.~Raiffa, G.L. Thompson, and R.M. Thrall.
\newblock The double description method.
\newblock In {\em Contributions to theory of games, Vol. 2}. Princeton
  University Press, New Jersey, Princeton, 1953.

\bibitem{kolmogorov2}
A.~N. Kolmogorov.
\newblock {\em Grundbegriffe der Wahrscheinlichkeitsrechnung}.
\newblock Springer, Berlin, 1933.
\newblock English translation in \cite{kolmogorov2e}.

\bibitem{cdd-pck}
Komei Fukuda.
\newblock {cdd} program.
\newblock 2000.

\bibitem{2000-poly}
Itamar Pitowsky and Karl Svozil.
\newblock New optimal tests of quantum nonlocality.
\newblock {\em Physical Review}, A, to appear, 2001.
\newblock {\tt arXiv:quant-ph/0011060}.

\bibitem{kolmogorov2e}
A.~N. Kolmogorov.
\newblock {\em Foundations of the Theory of Probability}.
\newblock Chelsea, New York, 1956.

\end{thebibliography}

\end{document}